\documentclass[prb,aps,twocolumn,groupedaddress,floats,showpacs,final,superscriptaddress]{revtex4-2}
\usepackage[latin3]{inputenc}
\usepackage[makeroom]{cancel}
\usepackage{graphicx}
\usepackage{amsmath}
\usepackage{amsfonts}
\usepackage{amssymb}
\usepackage{chngcntr}
\usepackage{color}
\usepackage{cancel}
\usepackage[left=2cm,right=2cm,top=2cm,bottom=2cm]{geometry}

\usepackage{graphicx}
\usepackage{dcolumn}
\usepackage{bm}
\usepackage{simplewick}
\usepackage{array}
\usepackage{appendix}

\RequirePackage[
   hyperindex,colorlinks,bookmarksnumbered,
   plainpages=true,pdfstartview=FitH]{hyperref}
\hypersetup{linkcolor=blue,urlcolor=blue,citecolor=blue}
\usepackage{hyperref}

\definecolor{purple}{rgb}{0.5,0,0.6}

\begin{document}

\title{Vertex corrections to nonlinear photoinduced currents in 2D superconductors}




\author{A.~V.~Parafilo}
\email[Corresponding author: ]{aparafil@ibs.re.kr}
\affiliation{Center for Theoretical Physics of Complex Systems, Institute for Basic Science (IBS), Daejeon 34126, Korea}



\author{V.~M.~Kovalev}
\affiliation{Rzhanov Institute of Semiconductor Physics, Siberian Branch\\ of Russian Academy of Science, Novosibirsk 630090, Russia}
\affiliation{Novosibirsk State Technical University, Novosibirsk 630073, Russia}

\author{I.~G.~Savenko}
\email[Corresponding author: ]{ivan.g.savenko@gmail.com}
\affiliation{Department of Physics, Guangdong Technion -- Israel Institute of Technology, 241 Daxue Road, Shantou, Guangdong, China, 515063}
\affiliation{Technion -- Israel Institute of Technology, 32000 Haifa, Israel}

\date{\today}

\begin{abstract}
The emergence of a rectified steady-state supercurrent as a response to the photoexcited current of the quasiparticles constitutes the concept of superconducting photodiode. This phenomenon occurs in a two-dimensional thin superconducting film with a built-in DC supercurrent, which is exposed to the circularly polarized external electromagnetic field. The flow of a Cooper-pair condensate, resulting as a second-order photo-response in a direction transverse to the initially built-in supercurrent, represents a superconducting counterpart to the photogalvanic effect. In this paper, we examine the photodiode supercurrent by restoring gauge invariance within the mean-field BCS framework. To achieve this, we derive an impurity-sensitive BCS-interaction-induced correction to the vertex function by performing self-consistent calculations within Keldysh Green's function technique. The resulting photodiode current can be utilized for spectroscopic analysis of typical relaxation times in superconducting films.

\end{abstract}

\maketitle

\section{Introduction}
Light-matter interaction in superconductors is a long-standing, broad and challenging research topic~\cite{RevModPhys.46.587, 10.1007/978-1-4684-1863-7_9, RevModPhys.77.721, dressel,  PhysRevLett.123.217004}. 
Immediately after discovering the BCS theory, it was shown that the conventional single-band superconductors do not absorb electromagnetic (EM) waves for symmetry reasons if disregarding the impurities~\cite{Tinkham, PhysRev.108.1175, Mahan}. 
The crucial aspect here is that the momentum-conserving photo-excitations of quasiparticles due to breaking of the Cooper pairs are forbidden in single-band superconductors due to combination of inversion and particle-hole symmetries~\cite{PhysRevB.95.014506,NagaosaOptRe2021, PhysRevB.106.214526,PhysRevB.106.094505,PhysRevB.106.L220504}. 
However, impurities lead to the lack of
the momentum conservation, thus lifting this obstacle~\cite{PhysRev.111.412, PhysRev.156.470, PhysRev.156.487,ZIMMERMANN199199, PhysRev.165.588}. 

To achieve a steady-state photo-induced electric current in a superconductor, 
one has to consider the nonlinear in the external EM field amplitude contributions, thus the direction of the current becomes time-independent.
However, the second-order effects (with respect to the external electric field strength) are usually weaker than the linear effects unless the linear effects vanish.
Another difficulty is that the mean-field BCS theory is not gauge-invariant by itself. 

Usually, the gauge invariance is restored by accounting for the BCS-interaction-induced vertices~\cite{PhysRev.117.648, Schrieffer}. 
This procedure is rather tricky, and it has only recently been applied for the nonlinear photoresponse in superconductors~\cite{PhysRevB.108.224516,nonlineargauge1,nonlineargauge}. In these works, the selection rule that prohibits optical absorption was overcome by considering the non-parabolic nature of the electron spectrum~\cite{PhysRevB.108.224516} or by examining a superconductor with multiple bands~\cite{nonlineargauge1,nonlineargauge} instead of focusing on impurities.

In this manuscript, we develop a theory of vertex-induced corrections to the nonlinear photoinduced electric current in the isotropic single-band 2D superconducting thin film.
We consider the case of an external circularly-polarized EM field ${\bf \mathcal{A}}(t)={\bf \mathcal{A}}\exp{(-i\omega t)}+{\bf \mathcal{A}}^*\exp{(i\omega t)}$ with $\mathcal{A}=\mathcal{A}_0(1, i\sigma)$, where $\sigma=\pm 1$ reflects left/right circular polarization, exerted normally to the 2D superconducting sample, and the case when inversion symmetry is broken by a built-in longitudinal supercurrent~\cite{ PhysRevLett.122.257001,PhysRevLett.125.097004}.

An important benefit of nonlinear response effects, 
is the gate-tunability, which can be used, e.g., for current rectification or light-control of the particle transport~\cite{RefNonlineHallEff01}. 
The superconducting (SC) analog of current rectification without an external electromagnetic source is realized in what is known as {\it a SC diode}. It is a device that allows a dissipationless current to flow in one direction under zero-bias voltage condition, while a resistive normal current flows in the opposite direction at finite voltages~\cite{RefScDiods2, RefStrambini, nadeem}. 
The SC diode effect occurs since the SC critical current has different values for currents of opposite directions, which is a consequence of both mirror and time-reversal symmetries breaking. 
Various origins of this rectification phenomenon in superconducting materials have been actively studied in the literature recently~\cite{PhysRevLett.128.037001,PhysRevLett.128.177001,He_2022,PhysRevLett.99.067004,sciadv.abo0309,PhysRevX.12.041013}.



In our recent studies~\cite{PhysRevB.108.L180509,PhysRevB.110.205413,Parafilo_2025}, we investigated the nonlinear photoinduced transport of quasiparticles in 2D superconductors, revealing two closely related effects: the {\it anomalous supercurrent Hall effect} and the {\it SC photodiode effect} -- characterized by the transverse charge transport in the absence of an external magnetic field. These two manifestations represent the same underlying phenomenon observed in different temperature regimes: at zero temperature ($T=0$) and near the superconducting critical temperature ($T\lesssim T_c$), respectively. We have shown that the simultaneous breaking of time-reversal and inversion symmetries, together with the impurity-induced disorder that breaks Galilean invariance, results in (i) current rectification and (ii) tunable supercurrent control via external EM drive. The former effect arises from the optical alignment of quasiparticle momenta induced by an oscillating electric field. For a circularly polarized electromagnetic field, the resulting phenomenological relation takes the form
${\bf j} \propto i[{\bf p}_s \times ({\bf \mathcal{A}} \times {\bf \mathcal{A}}^*)]$,
where ${\bf p}_s$ denotes the built-in momentum of Cooper pairs.
This effect is associated with an additional Cooper pair flow ($\mathbf{j}_{\rm sc}=-\textbf{j}$) that emerges to counteract charge accumulation along the transverse edges of the superconducting sample (similar to what is described in Refs.~\cite{PhysRevLett.125.097004, PhysRevB.18.5116, PhysRevB.21.1842,PhysRevB.65.064531}) and, correspondingly, to prevent the penetration of a static electric field in the superconductor. 

These findings open promising avenues for implementing non-dissipative, light-controlled elements in SC logical devices. Therefore, we employ the SC photodiode device proposed in~\cite{PhysRevB.108.L180509,Parafilo_2025} as a platform to investigate the vertex correction induced by BCS interactions.

\section{Microscopic theory}
The Hamiltonian of a single-band isotropic 2D $s$-wave superconductor under an external EM field reads ($\hbar=k_B=c=1$)
\begin{eqnarray}
\label{Hamiltonian}
\hat{H}=
\left(\begin{array}{cc}
\frac{\left[{\bf p}-{\bf p}_s-e\mathcal{A}(t)\right]^2}{2m}-\epsilon_F & \Delta \\
\Delta  & \epsilon_F-\frac{\left[{\bf p}+{\bf p}_s+e\mathcal{A}(t)\right]^2}{2m}
\end{array}\right),
\end{eqnarray}
where $\epsilon_F$ is the Fermi energy and $\Delta$ is a SC gap, which obeys the self-consistent equation
\begin{eqnarray}
1=\lambda \int \frac{d\textbf{p}}{(2\pi)^2}\frac{\tanh\left[\sqrt{\xi^2_{\textbf{p}}+\Delta^2}/2T\right]}{2\sqrt{\xi^2_{\textbf{p}}+\Delta^2}}.   
\end{eqnarray}
Here, $\xi_\textbf{p}=\textbf{p}^2/2m-\epsilon_F$ is the electron kinetic energy, and $\lambda$ is a strength of the BCS electron-electron interaction.
The electric current is
\begin{eqnarray}\label{Current}
{\bf j}(t)=-i\sum_{\textbf{p}}\,\textmd{Tr}\left\{\hat{{\bf j}}\,\hat{\mathcal{G}}^{<}(t,t)\right\},
\end{eqnarray}
where $\hat{\textbf{j}}=-\delta \hat{H}/\delta \mathcal{A}$,
and the `lesser' component of the Keldysh Green's function is determined by the equation $(i\partial_t-\hat{H})\hat{\mathcal{G}}(t, t')=\delta(t-t')$.
\begin{figure}
\includegraphics[width=1.\columnwidth]{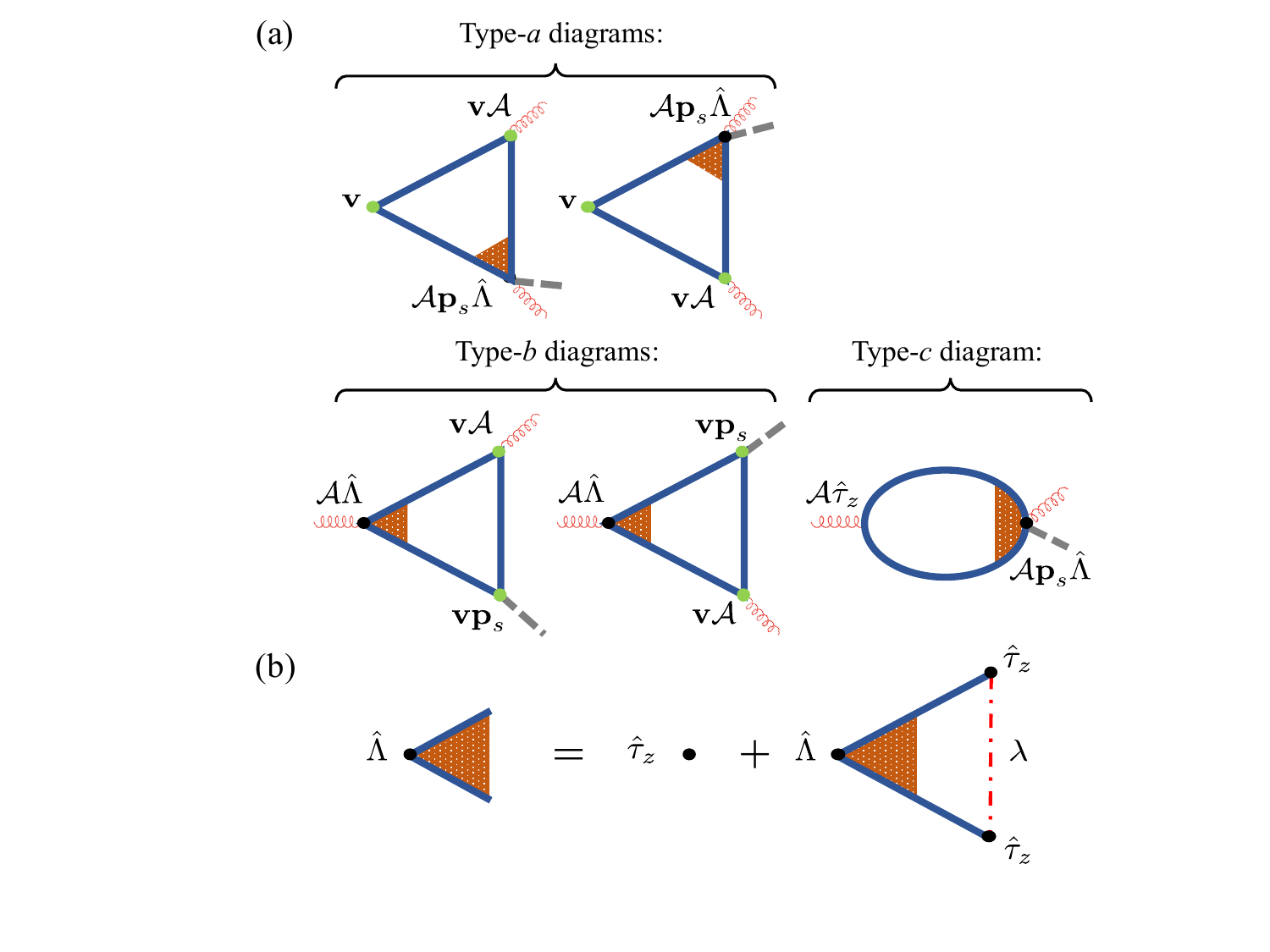} 
\caption{(a) Feynman diagrams providing the highest-order contribution to the photoinduced quasiparticle electric current. 
Blue lines are the Green's functions of quasiparticles, red wavy lines describe the external EM field $\mathcal{A}$; green dots stand for the quasiparticle velocity vertices $\textbf{v}$, and dashed lines show the supercurrent momentum ${\bf p}_s$. 
Orange triangles indicate the vertex renormalization $\hat \Lambda$ due to the BCS interaction. 
(b) Graphical illustration of the self-consistent equation for the vertex correction, see Eq.~(\ref{vertex}). 
Here, the red dash-dotted line indicates the BCS two-electron interaction of strength $\lambda$.}
\label{Fig1}
\end{figure}

We explore the stationary nonlinear photo-response by expanding $\hat{\mathcal{G}}^<(t,t')$ up to the first order with respect to $\textbf{p}_s$ and up to the second order with respect to $\mathcal{A}(t)$. 
This procedure provides twelve diagrams for the total current density~\cite{PhysRevB.108.L180509,Parafilo_2025}. 
However, the main contribution to the transverse component of the current density in the case of circularly polarized light is due to five diagrams, split into three types, which we call type-a, type-b and type-c (Fig.~\ref{Fig1}(a)).
The corresponding components of the current density read
%
\begin{eqnarray}\label{curr1}
&&\textbf{j}^{(a)}=ie^3\sum_{
\textbf{p}}\int dt_1\int dt_2
\\
\nonumber
&&~\times\left\{
\textrm{Tr}\left[\textbf{v}\hat{g}(t,t_1) \textbf{v}\cdot\mathcal{A}(t_1)\hat{g}(t_1,t_2)\textbf{v}_s\cdot\mathcal{A}(t_2)\hat{\Lambda}\,\hat{g}(t_2,t)\right]^<\right.\\
\nonumber
&&~\left.+\textrm{Tr}\left[\textbf{v}\hat{g}(t,t_1) \textbf{v}_s\cdot\mathcal{A}(t_1)\hat{\Lambda}\,\hat{g}(t_1,t_2)\textbf{v}\cdot\mathcal{A}(t_2)\hat{g}(t_2,t)\right]^<
\right\},
\end{eqnarray}
\begin{eqnarray}
&&\textbf{j}^{(b)}=ie^3\sum_{
\textbf{p}}\int dt_1\int dt_2
\\\nonumber
&&~\times\left\{\textrm{Tr}\left[\mathcal{A}(t)\hat{\Lambda}\,\hat{g}(t,t_1) \textbf{v}\cdot\mathcal{A}(t_1)\hat{g}(t_1,t_2)\textbf{v}\cdot\textbf{v}_s\hat{g}(t_2,t)\right]^<\right.\nonumber\\
&&~\left.+\textrm{Tr}\left[\mathcal{A}(t)\hat{\Lambda}\,\hat{g}(t,t_1) \textbf{v}\cdot\textbf{v}_s\hat{g}(t_1,t_2)\textbf{v}\cdot\mathcal{A}(t_2)\hat{g}(t_2,t)\right]^<\right\},\nonumber
\end{eqnarray}
\begin{eqnarray}
\label{curr2}
&&\textbf{j}^{(c)}=\frac{ie^3}{m}\sum_{
\textbf{p}}\int dt_1\\
\nonumber
&&~~~~~~\times\textrm{Tr}\left[\mathcal{A}(t)\hat{\tau}_z\hat{g}(t,t_1)\textbf{v}_s\cdot\mathcal{A}(t_1)\hat{\Lambda}\,\hat{g}(t_1,t)\right]^<,
\end{eqnarray}
where $\hat{g}(t,t')$ is a bare Keldysh Green's function, which represents a solution of the equation $(i\partial_t-\hat{H})\hat{g}(t, t')=\delta(t-t')$ with the Hamiltonian~\eqref{Hamiltonian} in which $\textbf{p}_s=0$ and $\mathcal{A}_0=0$;
$\hat \Lambda$ is a vertex correction induced by the two-electron BCS interaction (we discuss it more below). 

Let us for a moment forget about the vertex correction, thus taking $\hat \Lambda=\hat\tau_z$. and find the stationary photo-induced current density, using Eqs.~(\ref{curr1}) and (\ref{curr2}).
For that, we apply the Langreth theorem: 
\begin{eqnarray}
(\hat{g}_1\hat{g}_2\hat{g}_3)^<=\hat{g}_1^<\hat{g}_2^A\hat{g}_3^A+\hat{g}_1^R\hat{g}_2^<\hat{g}_3^A+\hat{g}_1^R\hat{g}_2^R\hat{g}_3^<,    
\end{eqnarray}
where the superscripts $R,A,<$ stand for the retarded, advanced and lesser Green's functions. 
The corresponding Green's functions in the energy-momentum representation read 
\begin{eqnarray}\label{GF}
&&\hat{g}^{R}_{\epsilon}=\frac{\epsilon \eta_{\epsilon}+\xi_{\bf p} \hat{\tau}_z+\Delta \eta_{\epsilon}\hat{\tau}_x}{\left(\epsilon-\epsilon_{\bf p}+ \frac{i}{2\tau_{\bf p}}\right)\left(\epsilon+\epsilon_{\bf p}+ \frac{i}{2\tau_{\bf p}}\right)}, \\
&&\hat{g}^{A}_{\epsilon}=\left(\hat{g}^{R}_{\epsilon}\right)^{\ast}\quad,\quad\hat{g}^<_{\epsilon}=-f_{\epsilon}\left[\hat{g}_{\epsilon}^R-\hat{g}_{\epsilon}^A\right],\label{GF2}
\end{eqnarray}
where $\epsilon_{\bf p}=\sqrt{\xi_{\bf p}+\Delta^2}$ is a quasiparticle dispersion, and $f_{\epsilon}=[\exp(\epsilon/T)+1]^{-1}$ is the Fermi-Dirac distribution function. 

In Eqs.~(\ref{GF}) and (\ref{GF2}), we account for the relaxation processes due to impurity scattering characterized by the time $\tau_i= [2\pi \nu_0 n u_0^2]^{-1}$ with $\nu_0$ the 2DEG density of states, $n$ the density of impurities, and $u_0$ the impurity  potential. 
This provides a factor
\begin{eqnarray}
\eta_{\epsilon}=1+\frac{i}{2\tau_i}\frac{\textrm{sign}\{\epsilon\}}{\sqrt{\epsilon^2-\Delta^2}}  
\end{eqnarray}
in the numerator of Eq.~(\ref{GF}), and an effective relaxation time
\begin{eqnarray}\label{totaltime}
\frac{1}{\tau_{\bf p}}=\frac{1}{\tau_i}\frac{|\xi_{\bf p}|}{\epsilon_{\bf p}}+\frac{1}{\tau_E}.
\end{eqnarray}
in the poles of Green's functions. 
Here, employing the parameter $\tau_E$, we phenomenologically account for the energy relaxation of the quasiparticles after they have been excited by the EM field. 

It should be noted, that there exist different relaxation processes that may take place in 2D superconductors. 
Usually, the efficiency of these processes depends on the properties of the SC sample and EM field~\cite{PhysRevB.101.134508, SMITH2020168105,Ovchinnlsaakyan} (see also \cite{PhysRevB.57.1147,PhysRevB.61.7108}). The inelastic energy relaxation time $\tau_E$ plays the dominant role in the case $\omega\ll\Delta$ and $T_c-T\ll T_c$~\cite{Ovchinnlsaakyan}.

Furthermore, the calculation of the current using Eqs.~(\ref{curr1}) and~(\ref{curr2}) with bare Green's functions~(\ref{GF}) and~(\ref{GF2}) in the limit $\tau_{\bf p}\rightarrow \infty$ gives zero current due to a selection rule dictated by the combination of electron-hole and inversion symmetries~\cite{PhysRevB.95.014506}. 
This demonstrates the importance of the  scattering on impurities for the optical absorption in superconductors.
Alternatively, it can be interpreted as an absence of a nontrivial optical response when the current operator commutes with the Hamiltonian of a single-band superconductor with parabolic dispersion. Accounting for the impurity scattering breaks the Galilean invariance and, thus, lifts this selection rule, resulting in a finite contribution to the current densities Eqs.~(\ref{curr1}) and (\ref{curr2}), see also Refs.~\cite{PhysRevB.108.L180509,Parafilo_2025}.

Let us return to the discussion of the BCS interaction-induced vertex correction $\hat \Lambda$, depicted in Fig.~\ref{Fig1}(b). 
It is known that the mean-field approach to the BCS theory violates the gauge invariance, or, more specifically, the Ward identity, which represents the continuity equation written in terms of the Green's functions and associated with the local charge conservation~\cite{PhysRevB.95.014506, PhysRev.117.648}.  
The Ward's identity should also be checked in the calculation of the photo response, as it has been addressed in a number of papers~\cite{PhysRevB.95.014506,PhysRevB.106.094505,PhysRevB.106.L220504}.
A way to restore the gauge invariance~\cite{Peskin} in SC materials was proposed by Nambu~\cite{PhysRev.117.648} (see also~\cite{Schrieffer}). 
The principal idea behind it is to account for the BCS electron-electron interaction-induced correction to the vertex function lines in a standard ladder approximation.
Accounting for the vertex correction ensures the charge conservation in the system (providing the correctness of the calculations). 

The vertex corrections for the linear
and second-order optical response in the case of impurity-free samples were reported in Refs.~\cite{PhysRevB.95.014506,PhysRevB.106.094505,PhysRevB.106.L220504} and~\cite{PhysRevB.108.224516,nonlineargauge}, respectively.
The self-consistent equation for $\hat \Lambda (\omega)$ reads [see Fig.~\ref{Fig1}(b)]:
\begin{eqnarray}\label{vertex}
\hat \Lambda (\pm\omega)=\hat \tau_z +i\lambda\sum_{\epsilon\textbf{p}'}(f_{\epsilon}-f_{\epsilon\pm\omega})\hat \tau_z \hat g^R_{\epsilon\pm\omega}\hat \Lambda (\omega) \hat g^A_{\epsilon}\hat \tau_z.~~
\end{eqnarray}
At zero temperature and in a clean SC sample (when both the impurity scattering and inelastic relaxation processes are absent ($\tau_{i},\tau_E\rightarrow \infty$), the solution of Eq.~(\ref{vertex}) coincides with the vertex correction found in Refs.~\cite{PhysRevB.95.014506, PhysRevB.106.L220504}. 
In particular, in the case of a parabolic dispersion, the vertex is $\hat\Lambda(\omega)\approx \hat \tau_z (1+\Lambda_z)+\hat\tau_y\Lambda_y$ (with $\Lambda_z\approx 0$ and $\Lambda_y\sim \Delta/\omega$, which does not depend on $\lambda$). 
Substituting of such a vortex in the equations gives a vanishing optical conductivity~\cite{PhysRevB.106.L220504}.

However, in our case, the vertex correction should be constructed on the Green's functions renormalized by the impurity scattering (not the bare Green's functions). 
Straightforward calculations using~(\ref{vertex}) and~(\ref{GF}),~(\ref{GF2}) in the limit $\omega\ll\Delta$ and $\omega\tau_i\lesssim 1$ give (see Appendix~\ref{AppendixA} for the details):
\begin{eqnarray}\label{vertex2}
\hat\Lambda(\omega) \approx \hat \tau_z \left(1+ \frac{ig}{2\omega \tau_i}\frac{1}{1-\frac{ig}{2\omega\tau_i}}\right),
\end{eqnarray}
where $g=\nu_0\lambda$ is a dimensionless BCS interaction constant. 
Thus, $\hat\Lambda$ in Eq.~(\ref{vertex2}) only possesses the $z$ component, $\hat \Lambda \propto \hat\tau_z(1+\Lambda_z)$, and it explicitly depends on $g$ (unlike it was in the regime considered in Refs.~\cite{PhysRevB.95.014506, PhysRevB.106.L220504}). 


Furthermore, substituting Eqs.~(\ref{GF}), (\ref{GF2}) and (\ref{vertex2}) in~(\ref{curr1}) and~(\ref{curr2}), and performing algebraic derivations presented in Appendix~\ref{AppendixB} yields: 
\begin{eqnarray}\label{expr1}
&&j_y^{(a)}+j_y^{(b)}\equiv j_{\rm y}=4j_0 \left(\frac{\Delta}{2T}\right)^2\mathcal{I}_1\left(\frac{\Delta}{2T}\right)\nonumber\\ 
&&~~~~~~\times\frac{(\omega\tau_E)^3}{\left[1+(\omega\tau_E)^2\right]^2}\frac{\tau_E}{\tau_i}\frac{1+\frac{g}{4}\frac{\tau_E}{\tau_i}[1-(\omega\tau_E)^2]}{(\omega\tau_E)^2+\frac{g^2}{4}\left(\frac{\tau_E}{\tau_i}\right)^2},
\end{eqnarray}
and
\begin{eqnarray}
&&~~~~~~~~~~j_y^{(c)}=j_I+j_{II}~~\textrm{with}\\
\label{expr2}
&&j_I= 2j_0\frac{\omega\tau_E}{1+(\omega\tau_E)^2}\mathcal{F}(g,\omega)\,\mathcal{I}_2\left(\frac{\Delta}{2T}\right),\\
\label{expr3}
&&j_{II}=\frac{j_0}{4} \frac{\omega\tau_E}{1+(\omega\tau_E)^2}
\left(\frac{\Delta}{2T}\right)^2
\frac{\mathcal{F}(g,\omega)}{(2T\tau_i)^{2}}
\mathcal{I}_3\left(\frac{\Delta}{2T}\right),
\end{eqnarray}
and $j_0=\sigma (e^3p_s\mathcal{A}_0^2/\hbar^2 m\pi)$, where we restored the dimension by accounting for $\hbar^2$;
\begin{eqnarray}
\mathcal{F}(g,\omega)=1-g\frac{g\left(\frac{\tau_E}{\tau_i}\right)^2+2\frac{\tau_E}{\tau_i}(\omega\tau_E)^2}{g^2\left(\frac{\tau_E}{\tau_i}\right)^2+4(\omega\tau_E)^2},    
\end{eqnarray}
and 
\begin{eqnarray}
\mathcal{I}_1(y)=\int_{y}^{\infty} \frac{dx}{x^2\cosh^2(x)}\,\,,\,\,\mathcal{I}_2(y)=\int_{y}^{\infty}\frac{dx\sqrt{1-(y/x)^2}}{\cosh^2(x)} ,\nonumber\\
\mathcal{I}_3(y)=\int_{y}^{\infty}\frac{dx}{x^3\cosh^2(x)} \frac{1}{\sqrt{x^2-y^2}}\,\,\,\,\,\,\,\,\,\,\,\,\,\,\,\,\,\,\,\,\,\,\,\,\nonumber
\end{eqnarray}
are dimensionless integrals.

Formulas~\eqref{expr1}-\eqref{expr3} describe the main contributions to the rectified photoexcited electric current density as the second-order response of a 2D superconductor, and constitute the main result of this manuscript. 
Note, that the contributions~\eqref{expr1}-\eqref{expr3} vanish as $\tau_i,\tau_E\rightarrow\infty$, as expected.
Indeed, the photoinduced quasiparticle transport should vanish in a clean sample, as required by the Galilean invariance.

In the concept of the SC photodiode~\cite{PhysRevB.108.L180509,Parafilo_2025}, a supercurrent appears in the transverse direction to counterbalance the photoinduced current of quasiparticles, preventing the charge accumulation at the transverse boundaries of the sample. 
The resulting photodiode supercurrent reads as $j_{\rm sc}=-j_p$ with $j_p\equiv\left(j_y^{(a)}+j_y^{(b)}+j_y^{(c)}\right)$. 
\begin{figure}[t!]
\includegraphics[width=1\columnwidth]{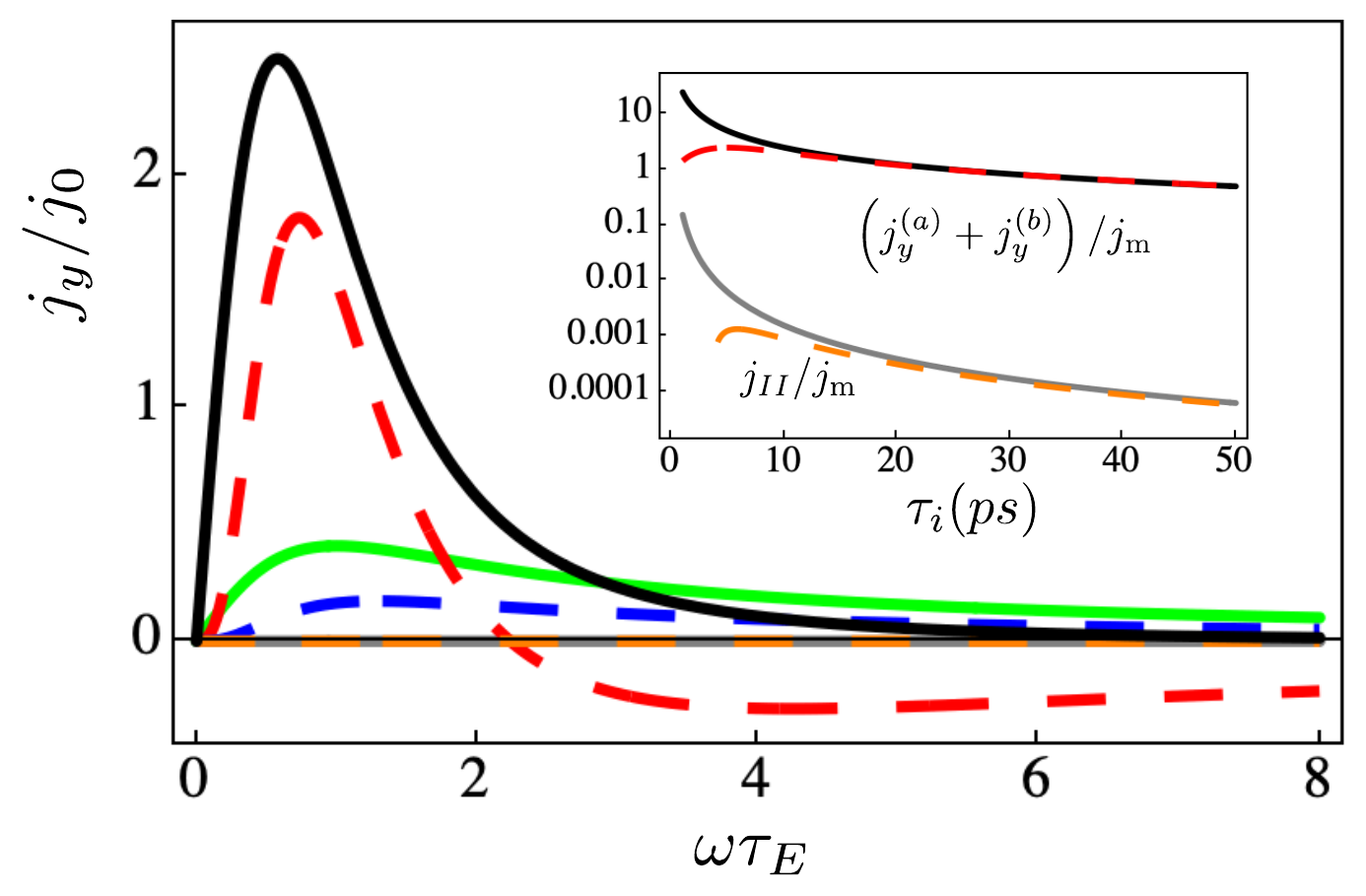}
\caption{The contributions to the electric current density Eqs.~(\ref{expr1})--(\ref{expr3}) as functions of normalized frequency $\omega\tau_E$. 
Black, green and gray solid curves show the contribution of $(j_y^{(a)}+j_y^{(b)})/j_0$, $j_I/j_0$, $j_{II}/j_0$ currents without the vertex correction ($g=0$), while red, blue and orange dashed curves corresponds to the same currents with $g=0.1$. 
Here, we used $T=9$ K, $T_c=10$ K ($\Delta_0\equiv \Delta(T=0)=1.5$ meV) and $\tau_i=10$ ps, $\tau_E=100$ ps. Inset: Maximum current densities $(j_y^{(a)}+j_y^{(b)})$ and $j_{II}/j_0$ as functions of impurity scattering time. Maximum for $j_{\rm y}$ current occurs at $\omega\tau_E\approx 1/\sqrt{3}$ (thus, $j_{m}=(3\sqrt{3}/4) j_0$), while for  $j_{II}$ current, the maximum is at $\omega\tau_E=1$ (with $j_m=j_0/8$). 
The color scheme is the same as in the main plot: solid curves indicate $g=0$ case and dashed curves correspond to $g=0.1$.
}
\label{Fig2}
\end{figure}
%
%
%


\section{Results and discussion} 
Figure~\ref{Fig2} shows the (normalized) spectrum of individual contributions described by Eqs.~(\ref{expr1})--(\ref{expr3}) to the photexcited current without and with the BCS--induced vertex correction. Solid black, green and gray curves stand for $j_y^{(a)}+j_y^{(b)}$,  $j_I$, and $j_{II}$ in the case $g=0$, respectively. Results Eqs.~(\ref{expr1})--(\ref{expr3}) at $g=0$ fully coincide with the equations obtained in Ref.~\cite{Parafilo_2025}. 

The main contribution to the transverse current density originates from the type-{\it a} and type-{\it b} diagrams since they are proportional to the ratio $\tau_E/\tau_i$, which can be rather large for typical relaxation times in  SC thin films~\cite{PhysRevB.102.054501}. 
In the meantime, the contribution of the current $j_{II}$ is proportional to $(\tau_i T_c)^{-2}$, see Eq.~(\ref{expr3}). This results in a negligible contribution of current Eq.~(\ref{curr2}) since $\Delta\tau_i \gg 1$ for the given set of parameters. 
However, it should be noted that the situation can be different in strongly disordered SC samples when $\Delta\tau_i\lesssim 1$.
This case is illustrated in the inset of Fig.~\ref{Fig2}: the contribution of the type-{\it c} diagram (more precisely, $j_{II}$ current) increases with decreasing $\tau_i$. 
Nevertheless, a more advanced treatment is required to adequately describe the case of strongly disordered superconductors ($\Delta \tau_i\ll 1$), which is beyond the scope of this manuscript.

The influence of the vertex correction on individual currents $j_{\rm y}$, $j_I$, and $j_{II}$ is shown in Fig.~\ref{Fig2} as dashed red, blue, and orange curves, respectively. 
Evidently, the vertex corrections give a weak suppression of the photodiode current strength and a modification of its frequency dependence. 

The main contribution to the photodiode current from the type-{\it a} and type-{\it b} diagrams acquires cubic dependence at small frequencies $j_{\rm y}(\omega\tau_E\ll1)/j_0\propto \omega^3$ instead of linear frequency dependence at the $g=0$ limit. 
Moreover, $j_{\rm y}$ current have the following asymptotic $j_{\rm y}(\omega\tau_E\gg1)/j_0\propto - g/\omega$ contrary to the limit $j_{\rm y}(\omega\tau_E\gg1)/j_0\sim \omega^{-3}$ at the case $g=0$ (compare black and red curves in Fig.~\ref{Fig2}). 
Thus, the account of the BCS correction results in that the current changes its sign on the opposite at certain frequency.

The influence of the vertex correction on the currents $j_I$ and $j_{II}$ consists only in the suppression of their strength with the increase of $g$. 
This remains true even when $\tau_i$ is reduced, see dashed curves in the inset of Fig.~\ref{Fig2}. 
Also note that at $g\geq 0.2$, the contribution of the type-{\it c} diagram can be completely neglected.

\begin{figure}
\includegraphics[width=1\columnwidth]{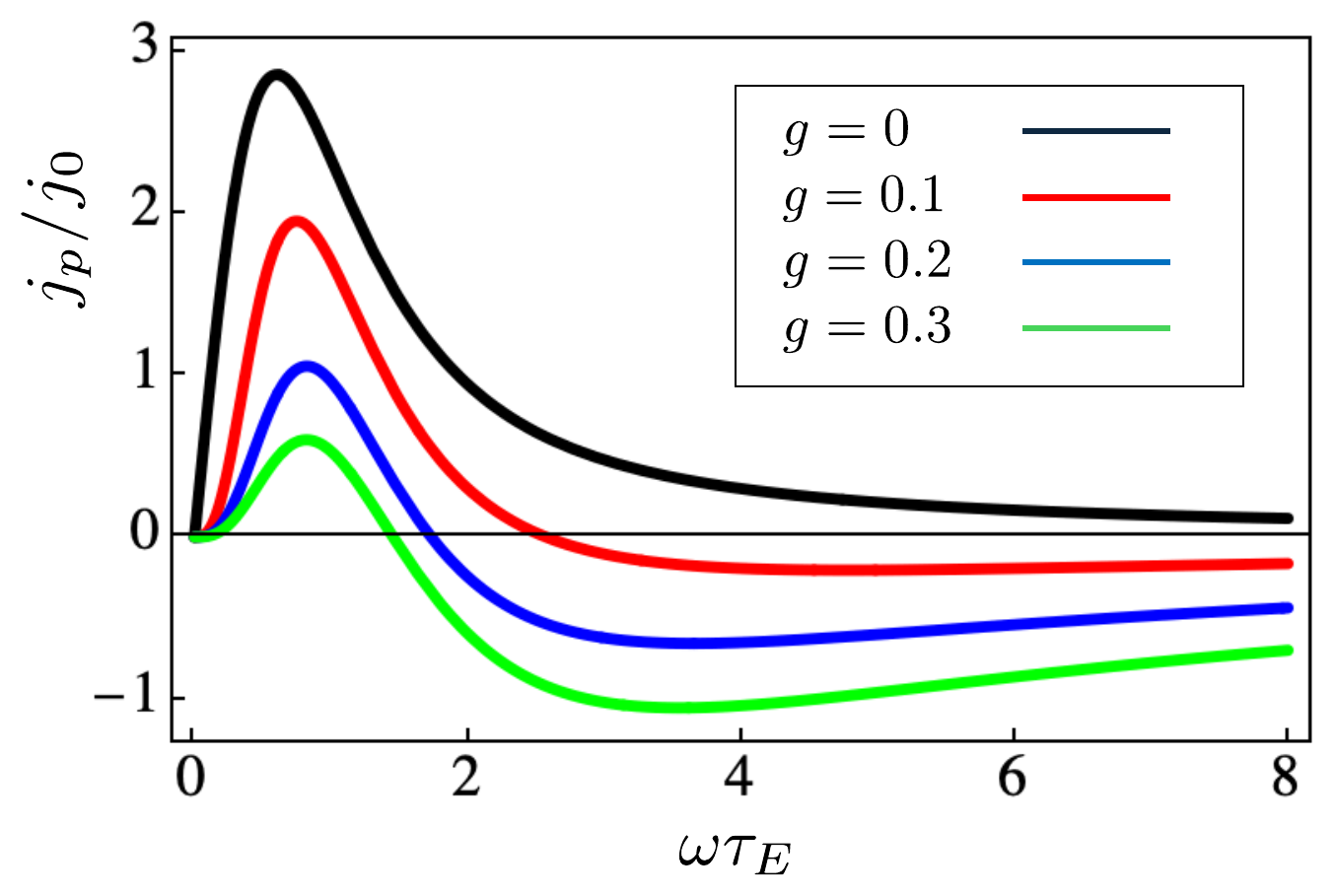}
\caption{Spectrum of the total photodiode current density $j_p=j_y^{(a)}+j_y^{(b)}+j_y^{(c)}$ normalzied by $j_0$ for different values of the BCS-induced vertex correction $g$. All parameters are the same as for Fig.~\ref{Fig2}. Note, current $j_I$ gives negligible contribution for $g\geq0.2$.
}
\label{Fig3}
\end{figure}

Figure~\ref{Fig3} demonstrates the total contribution of all the diagrams (type {\it a}-{\it c}) without (black) and with (other colors) the BCS-induced vertex correction. 
In the given set of parameters (when $\tau_i=10$~ps), the contribution of $j_{II}$ current is vanishingly small, while the contribution of $j_I$ current becomes negligible at $g\geq 0.2$. 

Next, let us estimate the magnitude of the photodiode current. 
For that, we can compare the induced current with the magnitude of the built-in longitudinal supercurrent. 
The momentum of the Cooper pair can be presented via the supercurrent as $p_s=j_sm/(en_s)$, where $n_s=n_s(T)$ is the density of particles in the condensate. 
Taking the density of the 2DEG as $n=0.6\cdot 10^{14}$~cm$^{-2}$, we find $n_s=6\cdot 10^{12}$~cm$^{-2}$ at $T=9$~K. 
Here, we used $\Delta_0\equiv \Delta(T=0)=17.65$~K ($\Delta_0=1.5$~meV) that gives $T_c=\Delta_0(\exp[0.577]/\pi)\approx 10$~K. 
Therefore, for the EM field intensity $I=1$~${\rm W/cm^2}$, the frequency $\omega=15$~GHz, and the relaxation times  $\tau_E=10^{-10}$~s and $\tau_i=10^{-11}$~s), we find $j_p/j_{\rm s}\approx0.5$ for $g=0.2$. 
Thus, we conclude that the transverse current can give a considerable correction to the built-in supercurrent.

The discovered specific dependence of the total current $j_{p}$ on the frequency of the EM field and the ratio $\tau_E/\tau_i$ can be used for the spectroscopy of relaxation times in SC films. 
The analysis is particularly simple for $g>0.2$ when only $j_y^{(a)}+j_y^{(b)}$ contribution determines the spectral behavior of the photocurrent. 

Moreover, even though the parameters $\tau_i$ and $\tau_E$ are generally unknown, we can find a useful relation between them from the zero-current condition: $\tau_E/\tau_i=(4/g)(\omega^2\tau^2_E-1)$. 
Here, $\omega$ and $T$ are experimentally controlled parameters, while $g$ can be estimated from the temperature dependence of the SC gap.

The second and the third required relations (for the maximum at $\omega\tau_E\approx 1$ and the minima at $\omega\tau_E\approx 3.5\div 4$) can be acquired from the condition of an extremum: $dj_{\rm y}/d(\omega\tau_E)=0$ (the resulting analytical formulas are too cumbersome to present them here). 
Utilizing these conditions relations (acquired from $j_{\rm y}=0$ and $dj_{\rm y}/d(\omega\tau_E)=0$) is sufficient to independently evaluate $\tau_E$ and $\tau_i$.



\section{Conclusions} 
We developed an analytical microscopic theory of light-matter interaction in 2D superconductors, accounting for the BCS interaction-induced vertex corrections in the framework of the Keldysh nonequilibrium diagram technique. 
We examined the electric current density, which emerges in a 2D superconductor with a built-in dc supercurrent as a consequence of photo absorption, assuming that the frequency of the external  circularly polarized EM field is $\omega\ll \Delta/\hbar$ (the regime $T_c-T\ll T_c$), such that the light is absorbed by thermally excited quasiparticles. 
In particular, we studied the spectrum of the rectified transverse (Hall-like) photoinduced electric current density and investigated its dependence on the effective impurity scattering time $\tau_i$, energy relaxation time $\tau_E$, and the constant of BCS interaction $g$. 
We showed that the account of the vertex corrections results in a valuable qualitative and quantitative modification of the spectrum. 
We propose the idea that the specific frequency dependence of the photoexcited current (spectrum) can be used for the experimental determination of the characteristic times $\tau_E$ and $\tau_i$. 
Finally, the account of the vertex correction restores the gauge invariance of the theory and leads to the satisfaction of the Ward's identity.


\section{Acknowledgements} 
We were supported by the Institute for Basic Science in Korea (Project No.~IBS-R024-D1), the National Natural Science Foundation of China, the Natural Science Foundation of Guangdong Province (China), the Foundation for the Advancement of Theoretical Physics and Mathematics ``BASIS'',
and Ministry of Science and Higher
Education of the RF (Project FSUN-2023-0006).


\appendix

\begin{widetext}


\section{BCS vertex correction}
\label{AppendixA}

In this Appendix, we show the details of the the BCS interaction-induced vertex correction calculations.

Using the Green's functions renormalized by the impurity scattering, Eqs.~(\ref{GF}) and (\ref{GF2}), one can write a self-consistent equation for the vertex (Fig.~\ref{Fig2}(b)):
\begin{eqnarray}\label{tra}
\hat{\Lambda}^-(\mp\omega)=\hat{\tau}_z+i\lambda\sum_{\epsilon\textbf{p}}(f_{\epsilon\mp\omega}-f_\epsilon)\hat{\tau}_z\hat{g}_\epsilon^R\hat{\Lambda}^-(\mp\omega) \hat{g}_{\epsilon\mp\omega}^A\hat{\tau}_z=\hat{\tau}_z+i\lambda\sum_{\bf p}\hat{\tau}_z\hat{\cal F}_{\epsilon,\epsilon\mp\omega}\hat{\tau}_z.
\end{eqnarray}
Here, index ``$-$'' in $\hat\Lambda^-$ indicates that the frequency dependence appears in the second Green's function. 
Below, we also use the index ``$+$'' to determine the vertex correction $\hat\Lambda^{+}(\mp\omega)$ built on $\hat g^R_{\epsilon\mp\omega}\hat g^A_{\epsilon}$ Green's functions. The function $\hat{\mathcal{F}}_{\epsilon,\epsilon-\omega}$ in Eq.~(\ref{tra}) reads as
\begin{eqnarray}
\nonumber
\hat{\cal F}_{\epsilon,\epsilon\mp\omega}&=&\int_{-\infty}^\infty \frac{d\epsilon}{2\pi}
 (f_{\epsilon\mp\omega}-f_\epsilon)
 \frac{(\epsilon\eta_{\epsilon}+\xi_{\bf p} \hat{\tau}_z+\Delta\eta_{\epsilon}\hat{\tau}_x)\hat{\Lambda}^-((\epsilon\mp\omega)\eta^{\ast}_{\epsilon\mp\omega}+\xi_{\bf p} \hat{\tau}_z+\Delta\eta^{\ast}_{\epsilon\mp\omega}\hat{\tau}_x)}{(\epsilon-\epsilon_{\bf p}+i \delta)(\epsilon+\epsilon_{\bf p}+i \delta)(\epsilon\mp\omega-\epsilon_{\bf p}-i \delta)(\epsilon\mp\omega+\epsilon_{\bf p}-i \delta)},
\end{eqnarray}
where we used the notation $\delta=(1/2\tau_i)(|\xi_{\bf p}|/\epsilon_{\bf p})$. 
Note, Eq.~(\ref{tra}), obtained with the bare Green's functions ($\delta=0$), coincides with the one obtained in~\cite{PhysRevB.95.014506}. 
Taking a contour integral over the energy at zero temperature yields
\begin{eqnarray}
\nonumber
\hat{\cal F}_{\epsilon,\epsilon\mp\omega}
    &=&\mp\frac{i}{2\omega\epsilon_{\bf p}}
    \left\{
    \frac{[(-\epsilon_{\bf p}+i\delta)\eta_{-\epsilon_{\bf p}}+\xi_{\bf p}\hat{\tau}_z+\Delta\eta_{-\epsilon_{\bf p}}\hat{\tau}_x]\hat{\Lambda}^-(\mp\omega)[(-\epsilon_{\bf p}+i\delta)\eta^{\ast}_{-\epsilon_{\bf p}}+\xi_{\bf p}\hat{\tau}_z+\Delta\eta^{\ast}_{-\epsilon_{\bf p}p}\hat{\tau}_x]}{-2\epsilon_{\bf p}+2i\delta}\right.\nonumber\\
   && ~~~~~~~~~~~\left.
    +
    \frac{[(-\epsilon_{\bf p}-i\delta)\eta_{-\epsilon_{\bf p}}+\xi_{\bf p}\hat{\tau}_z+\Delta\eta_{-\epsilon_{\bf p}}\hat{\tau}_x]\hat{\Lambda}^-(\mp\omega)[(-\epsilon_{\bf p}-i\delta)\eta^{\ast}_{-\epsilon_{\bf p}p}+\xi_{\bf p}\hat{\tau}_z+\Delta\eta^{\ast}_{-\epsilon_{\bf p}}\hat{\tau}_x]}{2\epsilon_{\bf p}+2i\delta}
    \right\}.
\end{eqnarray}
Here, we consider the regime $\omega\ll\Delta$ and neglect $\omega$ as compared with $\Delta$ and $\epsilon_{\bf p}$ in the spirit of Ref.~\cite{Parafilo_2025}.

Using the decomposition $\hat{\Lambda}^-(\mp\omega)=\Lambda_0\hat{1}+\Lambda_x\hat{\tau}_x+\Lambda_y\hat{\tau}_y+(\Lambda_z+1)\hat{\tau}_z$, we can rewrite Eq.~(\ref{tra}) in the matrix form for the case $\omega\tau_i\lesssim 1$:
\begin{eqnarray}
\label{self}
\left(
\begin{array}{cc}
\Lambda_0+\Lambda_z  &\Lambda_x -i\Lambda_y \\
    \Lambda_x+i\Lambda_y & \Lambda_0-\Lambda_z
\end{array}
\right)
\approx\mp\lambda\sum_{\bf p} \frac{\hat{M}}{8\omega\epsilon^3_{\bf p}}\quad,\quad    
\hat{M}=i\frac{|\xi_{\bf p}|}{\epsilon_{\bf p}\tau_i}
\left(
\begin{array}{cc}
   - 4\Delta^2(\Lambda_z+1)  & 4\xi_{\bf p}^2\Lambda_x-4i\epsilon_{\bf p}^2\Lambda_y  \\
4\xi^2_{\bf p}\Lambda_x+4i\epsilon_{\bf p}^2\Lambda_y & 4\Delta^2(\Lambda_z+1)
\end{array}
\right).
\end{eqnarray}
Solving the self-consistent equation in the matrix form, we find: $\Lambda_x\approx0$, $\Lambda_y\approx0$, and 
\begin{eqnarray}
\Lambda_z = \pm i\lambda\frac{4\Delta^2}{8\omega \tau_i}\sum_{\bf p} \frac{|\xi_{\bf p}|}{\epsilon_{\bf p}^4}(1+\Lambda_z)\quad\Rightarrow\quad
\Lambda_z = \pm i\lambda\frac{\Delta^2}{2\omega \tau_i}\sum_{\bf p} \frac{|\xi_{\bf p}|}{\epsilon_{\bf p}^4}\frac{1}{1\mp i\frac{\lambda\Delta^2}{2\omega\tau_i}\sum_{\bf q}\frac{|\xi_{\bf q}|}{\epsilon_{\bf q}^4}}.
\end{eqnarray}
Evaluating the integral gives
%
\begin{eqnarray}\label{vertcorr1}
\hat\Lambda^{-}(\mp\omega) \approx \left(1\pm \frac{ig}{2\omega \tau_i}\frac{1}{1\mp\frac{ig}{2\omega\tau_i}}\right)\hat\tau_z,
\end{eqnarray}
where $g=\nu_0\lambda$ is the dimensionless strength of the two-electron BCS interaction. 
The vertex correction is well justified in the limit $\omega\tau_i\lesssim 1$, thus, the vertex for the clean sample ($\omega\gg \tau_i^{-1}\rightarrow0$) cannot be obtained from Eq.~(\ref{vertcorr1}). Besides, after straightforward calculations, it is easy to show that $\hat\Lambda^+(\mp\omega)=\hat\Lambda^-(\pm\omega)$.

\section{Transverse photodiode current}
\label{AppendixB}

In this Appendix, we provide an explicit evaluation of the diagrams which contribute to the transverse component of the electric current density, depicted in Fig.~\ref{Fig1}(a).


\subsection{Type-a diagrams from Fig.~\ref{Fig1}(a)}

Choosing the direction of the given dc supercurrent as $\textbf{p}_s=(p_s,0)$, we find the following stationary current (from Eq.~(\ref{curr1})):
\begin{eqnarray}\label{vert1}
&&j_y^{(a)}/(j_0\pi)=
-\sum_{\epsilon\textbf{p}}\,v_y^2(f_{\epsilon}-f_{\epsilon+\omega})\textmd{Tr}\left\{\hat{g}^{R}_{\epsilon}\hat{g}_{\epsilon+\omega}^{A}\hat{\tau}_z\hat{g}_{\epsilon}^{A}\right\}-\sum_{\epsilon\textbf{p}}\,v_y^2(f_{\epsilon}-f_{\epsilon+\omega})\textmd{Tr}\left\{\hat{g}_{\epsilon}^{A}\hat{g}^{R}_{\epsilon}\hat\tau_z\hat{g}_{\epsilon+\omega}^{R}\right\}\nonumber\\
&&+\sum_{\epsilon\textbf{p}}\,v_y^2(f_{\epsilon}-f_{\epsilon+\omega})\textmd{Tr}\left\{\hat{g}^{R}_{\epsilon}\hat{g}_{\epsilon+\omega}^{R}\hat{\Lambda}^{+}(\omega)\hat{g}_{\epsilon}^{A}\right\}
+\sum_{\epsilon\textbf{p}}v_y^2(f_{\epsilon}-f_{\epsilon+\omega})\textmd{Tr}\left\{\hat{g}_{\epsilon}^{A}\hat{g}^{R}_{\epsilon}\hat{\Lambda}^{-}(\omega)\hat{g}_{\epsilon+\omega}^{A}\right\}-(\omega\rightarrow -\omega),
\end{eqnarray}
where $j_0=(e^3/m\pi)\sigma\mathcal{A}_0p_s$. 
Here, we consider only the stationary contribution to the transverse current (leaving the second-harmonic generation beyond the scope of this manuscript). 
Furthermore, we assume that the vertex function between two retarded Green's functions (two advanced Green's functions)  is $\hat g^R_{\epsilon}\hat\Lambda^-(\omega) \hat g^R_{\epsilon+\omega}=\hat \tau_z$ ($\hat g^A_{\epsilon}\hat\Lambda^-(\omega) \hat g^A_{\epsilon+\omega}=\hat \tau_z$) due to the integration over the energy 
in the definition of the vertex function, Eq.~(\ref{tra}).

Taking a trace and integrating over the energy in the first term in Eq.~(\ref{vert1}) gives
\begin{eqnarray}\label{wver1}
&&i\sum_{\bf p}v_y^2(f_{\epsilon_{\bf p}}-f_{\epsilon_{\bf p}+\omega})\frac{2\xi_p\left\{2\left(\epsilon_{\bf p}-\frac{i|\xi_{\bf p}|}{2\tau_i\epsilon_{\bf p}}\right)^2\eta_{-\epsilon_{\bf p}}+\xi_{\bf p}^2+\left[\left(\epsilon_{\bf p}-\frac{i|\xi_{\bf p}|}{2\tau_i\epsilon_{\bf p}}\right)^2+\Delta^2\right]|\eta_{\epsilon_{\bf p}}|^2+\Delta^2(|\eta_{\epsilon_{\bf p}}|^2-\eta_{-\epsilon_{\bf p}}^2)\right\}}{2\epsilon_{\bf p}(-\frac{i}{\tau_{\bf p}})(2\epsilon_{\bf p}-\frac{i}{\tau_{\bf p}})^2(\omega-\frac{i}{\tau_{\bf p}})}\nonumber\\
&&+i\sum_{\bf p}v_y^2(f_{-\epsilon_{\bf p}}-f_{-\epsilon_{\bf p}+\omega})\frac{2\xi_{\bf p}\left\{2\left(\epsilon_{\bf p}+\frac{i|\xi_{\bf p}|}{2\tau_i\epsilon_{\bf p}}\right)^2\eta_{\epsilon_{\bf p}}+\xi_{\bf p}^2+\left[\left(\epsilon_{\bf p}+\frac{i|\xi_{\bf p}|}{2\tau_i\epsilon_{\bf p}}\right)^2+\Delta^2\right]|\eta_{-\epsilon_{\bf p}}|^2+\Delta^2(|\eta_{-\epsilon_{\bf p}}|^2-\eta_{\epsilon_{\bf p}}^2)\right\}}{-2\epsilon_{\bf p}(-\frac{i}{\tau_{\bf p}})(2\epsilon_{\bf p}+\frac{i}{\tau_{\bf p}})^2(\omega-\frac{i}{\tau_{\bf p}})}.\nonumber\\
\end{eqnarray}
In the integration over energy, we used the following usual approximation: we accounted for the imaginary parts of the poles, $\epsilon_{1,2}=\pm\epsilon\pm (i/2\tau_{\bf p})$ with $\tau_p\rightarrow \tau_i\epsilon_{\bf p}/|\xi_{\bf p}|$, while disregarding the imaginary part when substituting the poles in $\eta_{\epsilon}$.

The second term in Eq.~(\ref{tra}) after the integration reads as
\begin{eqnarray}\label{wver2}
&&-i\sum_{\bf p}v_y^2(f_{\epsilon_{\bf p}}-f_{\epsilon_{\bf p}+\omega})\frac{2\xi_{\bf p}\left\{2(\epsilon_{\bf p}+\frac{i|\xi_{\bf p}|}{2\tau_i\epsilon_{\bf p}})^2\eta_{\epsilon_{\bf p}}+\xi_{\bf p}^2+[(\epsilon_{\bf p}+\frac{i|\xi_{\bf p}|}{2\tau_i\epsilon_{\bf p}})^2+\Delta^2]|\eta_{\epsilon_{\bf p}}|^2+\Delta^2(|\eta_{\epsilon_{\bf p}}|^2-\eta_{\epsilon_{\bf p}}^2)\right\}}{2\epsilon_{\bf p}(\frac{i}{\tau_{\bf p}})(2\epsilon_{\bf p}+\frac{i}{\tau_{\bf p}})^2(\omega+\frac{i}{\tau_{\bf p}})}\nonumber\\
&&-i\sum_{\bf p}v_y^2(f_{-\epsilon_{\bf p}}-f_{-\epsilon_{\bf p}+\omega})\frac{2\xi_{\bf p}\left\{2(\epsilon_{\bf p}-\frac{i|\xi_{\bf p}|}{2\tau_i\epsilon_{\bf p}})^2\eta_{-\epsilon_{\bf p}}+\xi_{\bf p}^2+[(\epsilon_{\bf p}-\frac{i|\xi_{\bf p}|}{2\tau_i\epsilon_{\bf p}})^2+\Delta^2]|\eta_{-\epsilon_{\bf p}}|^2+\Delta^2(|\eta_{-\epsilon_{\bf p}}|^2-\eta_{-\epsilon_{\bf p}}^2)\right\}}{-2\epsilon_{\bf p}(\frac{i}{\tau_{\bf p}})(2\epsilon_{\bf p}-\frac{i}{\tau_{\bf p}})^2(\omega+\frac{i}{\tau_{\bf p}})}.\nonumber\\
\end{eqnarray}

Furthermore, the third and the fourth terms from Eq.~(\ref{tra}) read:
\begin{eqnarray}\label{wwithver1}
&&i\sum_{\bf p}v_y^2(-\omega f'_{\epsilon_{\bf p}})\left[\frac{2\xi_{\bf p}\left\{(1+\Lambda_z^+(\omega))\left[2(\epsilon_{\bf p}+\frac{i|\xi_p|}{2\tau_i\epsilon_{\bf p}})^2\eta_{\epsilon_{\bf p}}+\xi_{\bf p}^2+[(\epsilon_{\bf p}+\frac{i|\xi_{\bf p}|}{2\tau_i\epsilon_{\bf p}})^2+\Delta^2]|\eta_{\epsilon_{\bf p}}|^2+\Delta^2(\eta_{\epsilon_{\bf p}}^2-|\eta_{\epsilon_{\bf p}}|^2)\right]\right\}}{2\epsilon_{\bf p}(\frac{i}{\tau_{\bf p}})(2\epsilon_{\bf p}+\frac{i}{\tau_{\bf p}})^2(\omega+\frac{i}{\tau_{\bf p}})}\right.\nonumber\\
&&\left.+\frac{2\xi_{\bf p}\left\{(1+\Lambda_z^+(\omega))\left[2(\epsilon_{\bf p}-\frac{i|\xi_{\bf p}|}{2\tau_i\epsilon_{\bf p}})^2\eta_{-\epsilon_{\bf p}}+\xi_{\bf p}^2+[(\epsilon_{\bf p}-\frac{i|\xi_{\bf p}|}{2\tau_i\epsilon_{\bf p}})^2+\Delta^2]|\eta_{\epsilon_{\bf p}}|^2+\Delta^2(\eta_{-\epsilon_{\bf p}}^2-|\eta_{\epsilon_{\bf p}}|^2)\right]\right\}}{-2\epsilon_{\bf p}(\frac{i}{\tau_{\bf p}})(2\epsilon_{\bf p}-\frac{i}{\tau_{\bf p}})^2(\omega+\frac{i}{\tau_{\bf p}})}\right]
\end{eqnarray}
and
\begin{eqnarray}\label{wwithver2}
&&i\sum_{\bf p}v_y^2\omega f'_{\epsilon_{\bf p}}\left[\frac{2\xi_{\bf p}\left\{(1+\Lambda_z^-(\omega))\left[2(\epsilon_{\bf p}-\frac{i|\xi_{\bf p}|}{2\tau_i\epsilon_{\bf p}})^2\eta_{-\epsilon_{\bf p}}+\xi_{\bf p}^2+[(\epsilon_{\bf p}-\frac{i|\xi_{\bf p}|}{2\tau_i\epsilon_{\bf p}})^2+\Delta^2]|\eta_{\epsilon_{\bf p}}|^2-\Delta^2(|\eta_{\epsilon_{\bf p}}|^2-\eta_{-\epsilon_{\bf p}}^2)\right]\right\}}{2\epsilon_{\bf p}(-\frac{i}{\tau_{\bf p}})(2\epsilon_{\bf p}-\frac{i}{\tau_{\bf p}})^2(\omega-\frac{i}{\tau_{\bf p}})}\right.\nonumber\\
&&\left.+\frac{2\xi_{\bf p}\left\{(1+\Lambda_z^-(\omega))\left[2(\epsilon_{\bf p}+\frac{i|\xi_{\bf p}|}{2\tau_i\epsilon_{\bf p}})^2\eta_{\epsilon_{\bf p}}+\xi_{\bf p}^2+[(\epsilon_{\bf p}+\frac{i|\xi_{\bf p}|}{2\tau_i\epsilon_{\bf p}})^2+\Delta^2]|\eta_{\epsilon_{\bf p}}|^2-\Delta^2(|\eta_{\epsilon_{\bf p}}|^2-\eta_{\epsilon_{\bf p}}^2)\right]\right\}}{-2\epsilon_{\bf p}(-\frac{i}{\tau_{\bf p}})(2\epsilon_{\bf p}+\frac{i}{\tau_{\bf p}})^2(\omega-\frac{i}{\tau_{\bf p}})}\right].
\end{eqnarray}
Here, we also simplified the difference between the distribution functions by expanding this difference over the small frequency, $\omega\ll\Delta$. 

Now, we can combine the terms~\eqref{wver2} and~\eqref{wwithver1}, yielding:
\begin{eqnarray}
&&i\frac{e^3}{m}\sigma \mathcal{A}_0^2p_s\sum_{\bf p} v_y^2(-\omega f'_{\epsilon_{\bf p}})\frac{\Delta^2}{\left(\omega +\frac{i}{\tau_{\bf p}}\right)}\frac{\xi_{\bf p}(1+\Lambda_z^+(\omega))}{2\epsilon_{\bf p}^3|\xi_{\bf p}|}\frac{2\tau_{\bf p}}{\tau_i},
\end{eqnarray}
as well as the terms Eqs.~(\ref{wver1}) and (\ref{wwithver2}):
\begin{eqnarray}
&&i\frac{e^3}{m}\sigma \mathcal{A}_0^2p_s\sum_{\bf p} v_y^2(-\omega f'_{\epsilon_{\bf p}})\frac{\Delta^2}{\left(\omega -\frac{i}{\tau_{\bf p}}\right)}\left\{-\frac{\xi_{\bf p}(1+\Lambda_z^-(\omega))}{2\epsilon_{\bf p}^3|\xi_{\bf p}|}\frac{2\tau_{\bf p}}{\tau_i}\right\}.
\end{eqnarray}

We can simplify $\Lambda_z^{\pm}(\omega)$ in the expressions above  assuming $g/\omega\tau_i\ll1$, yielding
\begin{eqnarray}\label{poi}
&&i\frac{e^3}{m}\sigma \mathcal{A}_0^2p_s\sum_{\bf p} v_y^2(- f'_{\epsilon_{\bf p}})\frac{\omega\Delta^2}{\left(\omega^2 +\frac{1}{\tau_{\bf p}^2}\right)}\frac{\xi_{\bf p}}{2\epsilon_{\bf p}^3|\xi_{\bf p}|}\frac{2\tau_{\bf p}}{\tau_i }\left(-\frac{2i}{\tau_{\bf p}} +\left(\omega-\frac{i}{\tau_{\bf p}}\right)\Lambda_z^+(\omega)-\left(\omega+\frac{i}{\tau_{\bf p}}\right)\Lambda^-_z(\omega)\right)\nonumber\\
=&&i\frac{e^3}{m}\sigma \mathcal{A}_0^2p_s\sum_{\bf p} v_y^2(- f'_{\epsilon_{\bf p}})\frac{\omega\Delta^2}{\left(\omega^2 +\frac{1}{\tau_{\bf p}^2}\right)}\frac{\xi_{\bf p}}{2\epsilon_{\bf p}^3|\xi_{\bf p}|}\frac{2\tau_{\bf p}}{\tau_i}\left(-\frac{2i}{\tau_{\bf p}} +\frac{ig}{\tau_i}+\frac{i g}{2\tau_{\bf p}(\omega\tau_i)^2}\right).
\end{eqnarray}

In the same way, we can find another term in Eq.~(\ref{vert1}) associated with the terms $\omega\rightarrow -\omega$.
The resulting expression is similar to Eq.~(\ref{poi}) with the replacement  $\{\omega,\sigma\}\rightarrow \{-\omega,-\sigma\}$.  

Combining two contributions provides
\begin{eqnarray}\label{vertexresult1}
j_y^{(a)}=\frac{2e^3}{m\pi}\sigma \mathcal{A}_0^2p_s\frac{\omega\Delta^2}{\left(\omega^2 +\tau_E^{-2}\right)}\int_{\Delta}^{\infty}d\left( \frac{\epsilon_{\bf p}}{2T}\right)\frac{1}{\cosh^2\left(\frac{\epsilon_{\bf p}}{2T}\right)}\frac{1}{\epsilon_{\bf p}^2}\frac{1}{\tau_i }\left(1 -g\frac{ \tau_E}{2 \tau_i}-\frac{ g^2}{4(\omega\tau_i)^2}\right),
\end{eqnarray}
where $\tau_{\bf p}=\tau_i\tau_E \varepsilon_{\bf p}/(\tau_E|\xi_{\bf p}|+\tau_i\varepsilon_{\bf p})\approx \tau_E$ since we consider the energies around the Fermi energy, $\xi_{\bf p}\rightarrow0$. 


\subsection{Type-b diagrams from Fig.~\ref{Fig1}(a)}

The expression for the current density, which correspond to type-b diagrams in Fig.~(\ref{Fig2}) reads as:
\begin{eqnarray}\label{eq88}
j_y^{(b)}=&&-\frac{e^3}{m}\mathcal{A}_0^2\sigma p_s\sum_{\epsilon\textbf{p}}v_x^2(f_{\epsilon+\omega}-f_{\epsilon})\textmd{Tr}\left\{\hat{g}^A_{\epsilon+\omega}\hat{g}^A_{\epsilon+\omega}\hat{\Lambda}^+(\omega)\,\hat{g}^R_{\epsilon}\right\}\nonumber\\&&-\frac{e^3}{m}\sigma \mathcal{A}_0^2p_s\sum_{\epsilon,\textbf{p}}v_x^2(f_{\epsilon+\omega}-f_{\epsilon})\textmd{Tr}\left\{\hat{g}^R_{\epsilon}\hat g^A_{\epsilon+\omega}\hat{\Lambda}^+(\omega)\hat{g}^R_{\epsilon}\right\}+(\omega,\sigma\rightarrow -\omega,-\sigma).
\end{eqnarray}
Note, that the vertex correction $\hat \Lambda^{+}(\omega)$ which appears between two Green's functions $\hat g^A_{\epsilon+\omega}\hat g^R_{\epsilon}$ coincides with the expression for $\hat\Lambda^+(\omega)$ found above for the vertex situated between $\hat g^R_{\epsilon+\omega}\hat g^A_{\epsilon}$. 

Taking the trace and integrating over the energy, for the first term in Eq.~(\ref{eq88}) we find
\begin{eqnarray}\label{nos}
&&-(j_0\pi)i\sum_{\textbf{p}}v_x^2\left[(f_{\epsilon_{\bf p}}-f_{\epsilon_{\bf p}-\omega})\frac{
2\xi_{\bf p}\left\{[(\epsilon_{\bf p}+\frac{i|\xi_{\bf p}|}{2\tau_i\epsilon_{\bf p}})^2+\Delta^2]\eta_{\epsilon_{\bf p}}^2+\xi_{\bf p}^2+2(\epsilon+\frac{i|\xi_{\bf p}|}{2\tau_i\epsilon_{\bf p}})^2|\eta_{\epsilon_{\bf p}}|^2\right\}(1+ \Lambda_z^+(\omega))}{2\epsilon_{\bf p}(2\epsilon_{\bf p}+\frac{i}{\tau_{\bf p}})^2(\omega-\frac{i}{\tau_{\bf p}})^2}\right.\nonumber\\
&&\left.+(f_{-\epsilon_{\bf p}}-f_{-\epsilon_{\bf p}-\omega})\frac{
2\xi_{\bf p}\left\{[(\epsilon_{\bf p}-\frac{i|\xi_{\bf p}|}{2\tau_i\epsilon_{\bf p}})^2+\Delta^2]\eta_{-\epsilon_{\bf p}}^2+\xi_{\bf p}^2+2(\epsilon-\frac{i|\xi_{\bf p}|}{2\tau_i\epsilon_{\bf p}})^2|\eta_{\epsilon_{\bf p}}|^2\right\}(1+\Lambda_z^+(\omega))}{-2\epsilon_{\bf p}(2\epsilon_{\bf p}-\frac{i}{\tau_{\bf p}})^2(\omega-\frac{i}{\tau_{\bf p}})^2}\right]\nonumber
\\&&=j_0\pi\sum_{\textbf{p}}v_x^2\frac{\omega f'_{\epsilon_{\bf p}}}{\epsilon_{\bf p}^3}\frac{\xi_{\bf p}}{|\xi_{\bf p}|}\frac{\Delta^2}{\tau_i}\frac{(1+\Lambda_z^+(\omega))}{(\omega-\frac{i}{\tau_{\bf p}})^2}.
\end{eqnarray}

The second term in Eq.~(\ref{eq88}) after the integration turns into
\begin{eqnarray}\label{nos2}
&&i(j_0\pi)\sum_{\bf p}\left[(f_{\epsilon_{\bf p}+\omega}-f_{\epsilon_{\bf p}})\frac{
2\xi_{\bf p}\left\{[(\epsilon_{\bf p}-\frac{i|\xi_{\bf p}|}{2\tau_i\epsilon_{\bf p}})^2+\Delta^2]\eta_{-\epsilon_{\bf p}}^2+\xi_{\bf p}^2+2(\epsilon-\frac{i|\xi_{\bf p}|}{2\tau_i\epsilon_{\bf p}})^2|\eta_{\epsilon_{\bf p}}|^2\right\}(1+\Lambda_z^+(\omega))}{2\epsilon_{\bf p}(2\epsilon_{\bf p}-\frac{i}{\tau_{\bf p}})^2(\omega-\frac{i}{\tau_{\bf p}})^2}\right.\nonumber\\
&&\left.+(f_{-\epsilon_{\bf p}+\omega}-f_{-\epsilon_{\bf p}})\frac{
2\xi_{\bf p}\left\{[(\epsilon_{\bf p}+\frac{i|\xi_{\bf p}|}{2\tau_i\epsilon_{\bf p}})^2+\Delta^2]\eta_{\epsilon_{\bf p}}^2+\xi_{\bf p}^2+2(\epsilon_{\bf p}+\frac{i|\xi_{\bf p}|}{2\tau_i\epsilon_{\bf p}})^2|\eta_{\epsilon_{\bf p}}|^2\right\}(1+\Lambda_z^+(\omega))}{-2\epsilon_{\bf p}(2\epsilon_{\bf p}+\frac{i}{\tau_{\bf p}})^2(\omega-\frac{i}{\tau_{\bf p}})^2}\right]\nonumber\\
&&=(j_0\pi)\sum_{\bf p}v_x^2\frac{\omega f'_{\epsilon_{\bf p}}}{\epsilon_{\bf p}^3}\frac{\xi_{\bf p}}{|\xi_{\bf p}|}\frac{\Delta^2}{\tau_i}\frac{(1+\Lambda_z^+(\omega))}{(\omega-\frac{i}{\tau_{\bf p}})^2}.
\end{eqnarray}

The other terms in Eq.~(\ref{eq88}) can be found by the replacement $\{\omega,\sigma\}\rightarrow \{-\omega,-\sigma\}$. 
Combining all the expressions yields the renormalized current density:
\begin{eqnarray}\label{vertexresult2}
j_y^{(b)}=\frac{2e^3}{m\pi}A_0^2\sigma p_s\Delta^2\frac{\tau_E}{\tau_i}\int_{\Delta/2T}^{\infty}\frac{d\epsilon_{\bf p}}{2T}\frac{(-1)}{\epsilon_{\bf p}^2\cosh^2\left(\frac{\epsilon_{\bf p}}{2T}\right)}\frac{\omega\tau_E}{\left(\omega^2\tau_E^2+1\right)^2}\left\{\left(\omega^2\tau_E^2-1\right)\left(1-\frac{g^2}{(2\omega\tau_i)^2}\right)+g\frac{\tau_E}{\tau_i}\right\}.
\end{eqnarray}

Since Eqs.~(\ref{vertexresult1}) and (\ref{vertexresult2}) provide the current of the same order of magnitude, we can combine both the expressions. 
The final expression for the type-a and type-b currents reads:
\begin{eqnarray}
j_y^{(a)}+j_y^{(b)}=4j_0\frac{\tau_E}{\tau_i}\left(\frac{\Delta}{2T}\right)^2\frac{\omega\tau_E}{(1+\omega^2\tau_E^2)^2}\int_{\Delta/2T}^{\infty}dx\frac{1}{x^2\cosh^2\left(x\right)}\left\{\left(1-\frac{g^2}{(2\omega\tau_i)^2}\right)+\frac{g}{4}\frac{\tau_E}{\tau_i}[1-(\omega\tau_E)^2]\right\}.
\end{eqnarray}

Here, we found a photoinduced current $j_y^{(a)}+j_y^{(b)}$ using the perturbation theory with respect to the parameter $g/\omega\tau_i$. 
However, the expression for the general case of arbitrary $g/\omega\tau_i$ can also be straightforwardly found (see Eq.~(\ref{expr1}) in the main text).
%
%
%

\subsection{Type-c diagram from Fig.~\ref{Fig1}}

The transverse component of the current density Eq.~(\ref{curr2}) reads as 
\begin{eqnarray}
j^{(c)}_y=-\frac{e^3}{m^2}\sigma\mathcal{A}_0^2p_s\sum_{\textbf{p}}\int\frac{d\epsilon}{2\pi}(f_{\epsilon+\omega}-f_{\epsilon})\frac{\textmd{Tr}\left\{\hat{\tau}_z[\epsilon\eta_{\epsilon}+\Delta \eta_{\epsilon}\hat{\tau}_x+\xi_{\textbf{p}}\hat{\tau}_z]\hat{\Lambda}^-(\omega)[(\epsilon+\omega)\eta_{\epsilon+\omega}^{\ast}+\Delta \eta_{\epsilon+\omega}^{\ast}\hat{\tau}_x+\xi_{\textbf{p}}\hat{\tau}_z]\right\}}{\left(\epsilon-\epsilon_{\textbf{p}}+\frac{i}{2\tau_{\textbf{p}}}\right)\left(\epsilon+\epsilon_{\textbf{p}}+\frac{i}{2\tau_{\textbf{p}}}\right)\left(\epsilon+\omega-\epsilon_{\textbf{p}}-\frac{i}{2\tau_{\textbf{p}}}\right)\left(\epsilon+\omega+\epsilon_{\textbf{p}}-\frac{i}{2\tau_{\textbf{p}}}\right)}\nonumber\\
+\frac{e^3}{m^2}\sigma\mathcal{A}_0^2p_s\sum_{\textbf{p}}\int\frac{d\epsilon}{2\pi}(f_{\epsilon-\omega}-f_{\epsilon})\frac{\textmd{Tr}\left\{\hat{\tau}_z[\epsilon\eta_{\epsilon}+\Delta \eta_{\epsilon}\hat{\tau}_x+\xi_{\textbf{p}}\hat{\tau}_z]\hat{\Lambda}^{-}(-\omega)[(\epsilon-\omega)\eta_{\epsilon-\omega}+\Delta \eta_{\epsilon-\omega}\hat{\tau}_x+\xi_{\textbf{p}}\hat{\tau}_z]\right\}}{\left(\epsilon-\epsilon_{\textbf{p}}+\frac{i}{2\tau_{\textbf{p}}}\right)\left(\epsilon+\epsilon_{\textbf{p}}+\frac{i}{2\tau_{\textbf{p}}}\right)\left(\epsilon-\omega-\epsilon_{\textbf{p}}-\frac{i}{2\tau_{\textbf{p}}}\right)\left(\epsilon-\omega+\epsilon_{\textbf{p}}-\frac{i}{2\tau_{\textbf{p}}}\right)},\nonumber\\
\end{eqnarray}
where we accounted for the vertex correction.

Integrating over the energy and tracing the matrices yields
\begin{eqnarray}
j^{(c)}_y=\frac{2ie^3}{m^2}\sigma\mathcal{A}_0^2p_s\sum_{\textbf{p}}(\omega f'_{\epsilon_{\bf p}})\frac{\left(2\xi_{\bf p}^2+\eta_{\epsilon_{\bf p}}\eta_{-\epsilon_{\bf p}}\left[\left(\epsilon_{\bf p}-\frac{i|\xi_{\bf p}|}{2\tau_i\epsilon_{\bf p}}\right)^2-\Delta^2\right]+\eta_{-\epsilon_{\bf p}}\eta_{\epsilon_{\bf p}}\left[\left(\epsilon_{\bf p}+\frac{i|\xi_{\bf p}|}{2\tau_i\epsilon_{\bf p}}\right)^2-\Delta^2\right]\right)(1+\Lambda^-_z(\omega))}{\left(2\epsilon_{\bf p}\right)^2\left(\omega-\frac{i}{\tau_{\bf p}}\right)}\nonumber\\
-\frac{2ie^3}{m^2}\sigma\mathcal{A}_0^2p_s\sum_{\textbf{p}}(\omega f'_{\epsilon_{\bf p}})\frac{\left(2\xi_{\bf p}^2+\eta_{\epsilon_{\bf p}}\eta_{-\epsilon_{\bf p}}\left[\left(\epsilon_{\bf p}-\frac{i|\xi_{\bf p}|}{2\tau_i\epsilon_{\bf p}}\right)^2-\Delta^2\right]+\eta_{-\epsilon_{\bf p}}\eta_{\epsilon_{\bf p}}\left[\left(\epsilon_{\bf p}+\frac{i|\xi_{\bf p}|}{2\tau_i\epsilon_{\bf p}}\right)^2-\Delta^2\right]\right)(1+\Lambda^-_z(-\omega))}{\left(2\epsilon_{\bf p}\right)^2\left(\omega+\frac{i}{\tau_{\bf p}}\right)}.\nonumber\\\label{num}
\end{eqnarray}

The numerator in each term in Eq.~(\ref{num}) equals
\begin{eqnarray}
2\xi_{\bf p}^2+2\left(1+\frac{1}{4\tau_i^2 \xi_{\bf p}^2}\right)\left(\epsilon_{\bf p}^2-\Delta^2-\frac{\xi_{\bf p}^2}{4\tau_i^2\epsilon_{\bf p}^2}\right)=4\xi_{\bf p}^2+\frac{2\xi^2_{\bf p}}{4\tau_i^2}\frac{\Delta^2}{\epsilon_{\bf p}^2\xi^2_{\bf p}}.
\end{eqnarray}
In the mean time, 
\begin{eqnarray}
\frac{1}{\omega-\frac{i}{\tau_{\bf p}}}-\frac{1}{\omega+\frac{i}{\tau_{\bf p}}}=\frac{2i}{\tau_{\bf p}}\frac{1}{\omega^2+\frac{1}{\tau_{\bf p}^2}}.
\end{eqnarray}
After algebraic calculations, we find:
\begin{eqnarray}\label{int}
j^{(c)}_y=-\frac{e^3}{m^2}\sigma\mathcal{A}_0^2p_s\sum_{\textbf{p}}\frac{\omega f'_{\epsilon_{\bf p}}}{\omega^2+\frac{1}{\tau_E^2}}\frac{1}{\tau_E\epsilon_{\bf p}^2}\left[4\xi_{\bf p}^2+\frac{\Delta^2}{2\tau_i^2\epsilon_{\bf p}^2}\right]\left(1-\frac{g^2}{(2\omega\tau_i)^2}-\frac{g}{2}\frac{\tau_E}{\tau_i}\right).
\end{eqnarray}

Interestingly, Eq.~(\ref{int}) contains a term (the first term in the brackets) that does not depend on the impurity scattering time. 
Here, the phenomenologically introduced inelastic processes characterized by the effective time $\tau_E$ are responsible for breaking the Galilean invariance. 
In the limit $\tau_E\rightarrow \infty$ (for clean samples), the current density vanishes. 


\end{widetext}

\bibliography{biblio}

\begin{thebibliography}{48}%
\makeatletter
\providecommand \@ifxundefined [1]{%
 \@ifx{#1\undefined}
}%
\providecommand \@ifnum [1]{%
 \ifnum #1\expandafter \@firstoftwo
 \else \expandafter \@secondoftwo
 \fi
}%
\providecommand \@ifx [1]{%
 \ifx #1\expandafter \@firstoftwo
 \else \expandafter \@secondoftwo
 \fi
}%
\providecommand \natexlab [1]{#1}%
\providecommand \enquote  [1]{``#1''}%
\providecommand \bibnamefont  [1]{#1}%
\providecommand \bibfnamefont [1]{#1}%
\providecommand \citenamefont [1]{#1}%
\providecommand \href@noop [0]{\@secondoftwo}%
\providecommand \href [0]{\begingroup \@sanitize@url \@href}%
\providecommand \@href[1]{\@@startlink{#1}\@@href}%
\providecommand \@@href[1]{\endgroup#1\@@endlink}%
\providecommand \@sanitize@url [0]{\catcode `\\12\catcode `\$12\catcode
  `\&12\catcode `\#12\catcode `\^12\catcode `\_12\catcode `\%12\relax}%
\providecommand \@@startlink[1]{}%
\providecommand \@@endlink[0]{}%
\providecommand \url  [0]{\begingroup\@sanitize@url \@url }%
\providecommand \@url [1]{\endgroup\@href {#1}{\urlprefix }}%
\providecommand \urlprefix  [0]{URL }%
\providecommand \Eprint [0]{\href }%
\providecommand \doibase [0]{https://doi.org/}%
\providecommand \selectlanguage [0]{\@gobble}%
\providecommand \bibinfo  [0]{\@secondoftwo}%
\providecommand \bibfield  [0]{\@secondoftwo}%
\providecommand \translation [1]{[#1]}%
\providecommand \BibitemOpen [0]{}%
\providecommand \bibitemStop [0]{}%
\providecommand \bibitemNoStop [0]{.\EOS\space}%
\providecommand \EOS [0]{\spacefactor3000\relax}%
\providecommand \BibitemShut  [1]{\csname bibitem#1\endcsname}%
\let\auto@bib@innerbib\@empty
\bibitem [{\citenamefont {Tinkham}(1974)}]{RevModPhys.46.587}%
  \BibitemOpen
  \bibfield  {author} {\bibinfo {author} {\bibfnamefont {M.}~\bibnamefont
  {Tinkham}},\ }\href {https://doi.org/10.1103/RevModPhys.46.587} {\bibfield
  {journal} {\bibinfo  {journal} {Rev. Mod. Phys.}\ }\textbf {\bibinfo {volume}
  {46}},\ \bibinfo {pages} {587} (\bibinfo {year} {1974})}\BibitemShut
  {NoStop}%
\bibitem [{\citenamefont {Tinkham}(1970)}]{10.1007/978-1-4684-1863-7_9}%
  \BibitemOpen
  \bibfield  {author} {\bibinfo {author} {\bibfnamefont {M.}~\bibnamefont
  {Tinkham}},\ }in\ \href@noop {} {\emph {\bibinfo {booktitle} {Far-Infrared
  Properties of Solids}}}\ (\bibinfo  {publisher} {Springer US},\ \bibinfo
  {address} {Boston, MA},\ \bibinfo {year} {1970})\ pp.\ \bibinfo {pages}
  {223--246}\BibitemShut {NoStop}%
\bibitem [{\citenamefont {Basov}\ and\ \citenamefont
  {Timusk}(2005)}]{RevModPhys.77.721}%
  \BibitemOpen
  \bibfield  {author} {\bibinfo {author} {\bibfnamefont {D.~N.}\ \bibnamefont
  {Basov}}\ and\ \bibinfo {author} {\bibfnamefont {T.}~\bibnamefont {Timusk}},\
  }\href {https://doi.org/10.1103/RevModPhys.77.721} {\bibfield  {journal}
  {\bibinfo  {journal} {Rev. Mod. Phys.}\ }\textbf {\bibinfo {volume} {77}},\
  \bibinfo {pages} {721} (\bibinfo {year} {2005})}\BibitemShut {NoStop}%
\bibitem [{\citenamefont {Dressel}(2013)}]{dressel}%
  \BibitemOpen
  \bibfield  {author} {\bibinfo {author} {\bibfnamefont {M.}~\bibnamefont
  {Dressel}},\ }\href {https://doi.org/10.1155/2013/104379} {\bibfield
  {journal} {\bibinfo  {journal} {Adv. in Cond. Matt. Phys.}\ }\textbf
  {\bibinfo {volume} {2013}},\ \bibinfo {pages} {104379} (\bibinfo {year}
  {2013})}\BibitemShut {NoStop}%
\bibitem [{\citenamefont {Yin}\ \emph {et~al.}(2019)\citenamefont {Yin},
  \citenamefont {Zhang}, \citenamefont {Dai}, \citenamefont {Zhao},
  \citenamefont {Kreisel}, \citenamefont {Macam}, \citenamefont {Wu},
  \citenamefont {Miao}, \citenamefont {Huang}, \citenamefont {Martiny},\ and\
  \citenamefont {et.al.}}]{PhysRevLett.123.217004}%
  \BibitemOpen
  \bibfield  {author} {\bibinfo {author} {\bibfnamefont {J.-X.}\ \bibnamefont
  {Yin}}, \bibinfo {author} {\bibfnamefont {S.~S.}\ \bibnamefont {Zhang}},
  \bibinfo {author} {\bibfnamefont {G.}~\bibnamefont {Dai}}, \bibinfo {author}
  {\bibfnamefont {Y.}~\bibnamefont {Zhao}}, \bibinfo {author} {\bibfnamefont
  {A.}~\bibnamefont {Kreisel}}, \bibinfo {author} {\bibfnamefont
  {G.}~\bibnamefont {Macam}}, \bibinfo {author} {\bibfnamefont
  {X.}~\bibnamefont {Wu}}, \bibinfo {author} {\bibfnamefont {H.}~\bibnamefont
  {Miao}}, \bibinfo {author} {\bibfnamefont {Z.-Q.}\ \bibnamefont {Huang}},
  \bibinfo {author} {\bibfnamefont {J.~H.~J.}\ \bibnamefont {Martiny}},\ and\
  \bibinfo {author} {\bibnamefont {et.al.}},\ }\href
  {https://doi.org/10.1103/PhysRevLett.123.217004} {\bibfield  {journal}
  {\bibinfo  {journal} {Phys. Rev. Lett.}\ }\textbf {\bibinfo {volume} {123}},\
  \bibinfo {pages} {217004} (\bibinfo {year} {2019})}\BibitemShut {NoStop}%
\bibitem [{\citenamefont {Tinkham}(2004)}]{Tinkham}%
  \BibitemOpen
  \bibfield  {author} {\bibinfo {author} {\bibfnamefont {M.}~\bibnamefont
  {Tinkham}},\ }\href {http://www.worldcat.org/isbn/0486435032} {\emph
  {\bibinfo {title} {Introduction to Superconductivity}}},\ \bibinfo {edition}
  {2nd}\ ed.\ (\bibinfo  {publisher} {Dover Publications},\ \bibinfo {year}
  {2004})\BibitemShut {NoStop}%
\bibitem [{\citenamefont {Bardeen}\ \emph {et~al.}(1957)\citenamefont
  {Bardeen}, \citenamefont {Cooper},\ and\ \citenamefont
  {Schrieffer}}]{PhysRev.108.1175}%
  \BibitemOpen
  \bibfield  {author} {\bibinfo {author} {\bibfnamefont {J.}~\bibnamefont
  {Bardeen}}, \bibinfo {author} {\bibfnamefont {L.~N.}\ \bibnamefont
  {Cooper}},\ and\ \bibinfo {author} {\bibfnamefont {J.~R.}\ \bibnamefont
  {Schrieffer}},\ }\href {https://doi.org/10.1103/PhysRev.108.1175} {\bibfield
  {journal} {\bibinfo  {journal} {Phys. Rev.}\ }\textbf {\bibinfo {volume}
  {108}},\ \bibinfo {pages} {1175} (\bibinfo {year} {1957})}\BibitemShut
  {NoStop}%
\bibitem [{\citenamefont {Mahan}(1990)}]{Mahan}%
  \BibitemOpen
  \bibfield  {author} {\bibinfo {author} {\bibfnamefont {G.~D.}\ \bibnamefont
  {Mahan}},\ }\href@noop {} {\emph {\bibinfo {title} {{Many-Particle
  Physics}}}}\ (\bibinfo  {publisher} {Plenum Press, New York},\ \bibinfo
  {year} {1990})\BibitemShut {NoStop}%
\bibitem [{\citenamefont {Dai}\ and\ \citenamefont
  {Lee}(2017)}]{PhysRevB.95.014506}%
  \BibitemOpen
  \bibfield  {author} {\bibinfo {author} {\bibfnamefont {Z.}~\bibnamefont
  {Dai}}\ and\ \bibinfo {author} {\bibfnamefont {P.~A.}\ \bibnamefont {Lee}},\
  }\href {https://doi.org/10.1103/PhysRevB.95.014506} {\bibfield  {journal}
  {\bibinfo  {journal} {Phys. Rev. B}\ }\textbf {\bibinfo {volume} {95}},\
  \bibinfo {pages} {014506} (\bibinfo {year} {2017})}\BibitemShut {NoStop}%
\bibitem [{\citenamefont {Ahn}\ and\ \citenamefont
  {Nagaosa}(2021)}]{NagaosaOptRe2021}%
  \BibitemOpen
  \bibfield  {author} {\bibinfo {author} {\bibfnamefont {J.}~\bibnamefont
  {Ahn}}\ and\ \bibinfo {author} {\bibfnamefont {N.}~\bibnamefont {Nagaosa}},\
  }\href {https://doi.org/10.1038/s41467-021-21905-x} {\bibfield  {journal}
  {\bibinfo  {journal} {Nature Communications}\ }\textbf {\bibinfo {volume}
  {12}},\ \bibinfo {pages} {1617} (\bibinfo {year} {2021})}\BibitemShut
  {NoStop}%
\bibitem [{\citenamefont {Crowley}\ and\ \citenamefont
  {Fu}(2022)}]{PhysRevB.106.214526}%
  \BibitemOpen
  \bibfield  {author} {\bibinfo {author} {\bibfnamefont {P.~J.~D.}\
  \bibnamefont {Crowley}}\ and\ \bibinfo {author} {\bibfnamefont
  {L.}~\bibnamefont {Fu}},\ }\href
  {https://doi.org/10.1103/PhysRevB.106.214526} {\bibfield  {journal} {\bibinfo
   {journal} {Phys. Rev. B}\ }\textbf {\bibinfo {volume} {106}},\ \bibinfo
  {pages} {214526} (\bibinfo {year} {2022})}\BibitemShut {NoStop}%
\bibitem [{\citenamefont {Radkevich}\ and\ \citenamefont
  {Semenov}(2022)}]{PhysRevB.106.094505}%
  \BibitemOpen
  \bibfield  {author} {\bibinfo {author} {\bibfnamefont {A.~A.}\ \bibnamefont
  {Radkevich}}\ and\ \bibinfo {author} {\bibfnamefont {A.~G.}\ \bibnamefont
  {Semenov}},\ }\href {https://doi.org/10.1103/PhysRevB.106.094505} {\bibfield
  {journal} {\bibinfo  {journal} {Phys. Rev. B}\ }\textbf {\bibinfo {volume}
  {106}},\ \bibinfo {pages} {094505} (\bibinfo {year} {2022})}\BibitemShut
  {NoStop}%
\bibitem [{\citenamefont {Papaj}\ and\ \citenamefont
  {Moore}(2022)}]{PhysRevB.106.L220504}%
  \BibitemOpen
  \bibfield  {author} {\bibinfo {author} {\bibfnamefont {M.}~\bibnamefont
  {Papaj}}\ and\ \bibinfo {author} {\bibfnamefont {J.~E.}\ \bibnamefont
  {Moore}},\ }\href {https://doi.org/10.1103/PhysRevB.106.L220504} {\bibfield
  {journal} {\bibinfo  {journal} {Phys. Rev. B}\ }\textbf {\bibinfo {volume}
  {106}},\ \bibinfo {pages} {L220504} (\bibinfo {year} {2022})}\BibitemShut
  {NoStop}%
\bibitem [{\citenamefont {Mattis}\ and\ \citenamefont
  {Bardeen}(1958)}]{PhysRev.111.412}%
  \BibitemOpen
  \bibfield  {author} {\bibinfo {author} {\bibfnamefont {D.~C.}\ \bibnamefont
  {Mattis}}\ and\ \bibinfo {author} {\bibfnamefont {J.}~\bibnamefont
  {Bardeen}},\ }\href {https://doi.org/10.1103/PhysRev.111.412} {\bibfield
  {journal} {\bibinfo  {journal} {Phys. Rev.}\ }\textbf {\bibinfo {volume}
  {111}},\ \bibinfo {pages} {412} (\bibinfo {year} {1958})}\BibitemShut
  {NoStop}%
\bibitem [{\citenamefont {Nam}(1967{\natexlab{a}})}]{PhysRev.156.470}%
  \BibitemOpen
  \bibfield  {author} {\bibinfo {author} {\bibfnamefont {S.~B.}\ \bibnamefont
  {Nam}},\ }\href {https://doi.org/10.1103/PhysRev.156.470} {\bibfield
  {journal} {\bibinfo  {journal} {Phys. Rev.}\ }\textbf {\bibinfo {volume}
  {156}},\ \bibinfo {pages} {470} (\bibinfo {year}
  {1967}{\natexlab{a}})}\BibitemShut {NoStop}%
\bibitem [{\citenamefont {Nam}(1967{\natexlab{b}})}]{PhysRev.156.487}%
  \BibitemOpen
  \bibfield  {author} {\bibinfo {author} {\bibfnamefont {S.~B.}\ \bibnamefont
  {Nam}},\ }\href {https://doi.org/10.1103/PhysRev.156.487} {\bibfield
  {journal} {\bibinfo  {journal} {Phys. Rev.}\ }\textbf {\bibinfo {volume}
  {156}},\ \bibinfo {pages} {487} (\bibinfo {year}
  {1967}{\natexlab{b}})}\BibitemShut {NoStop}%
\bibitem [{\citenamefont {Zimmermann}\ \emph {et~al.}(1991)\citenamefont
  {Zimmermann}, \citenamefont {Brandt}, \citenamefont {Bauer}, \citenamefont
  {Seider},\ and\ \citenamefont {Genzel}}]{ZIMMERMANN199199}%
  \BibitemOpen
  \bibfield  {author} {\bibinfo {author} {\bibfnamefont {W.}~\bibnamefont
  {Zimmermann}}, \bibinfo {author} {\bibfnamefont {E.}~\bibnamefont {Brandt}},
  \bibinfo {author} {\bibfnamefont {M.}~\bibnamefont {Bauer}}, \bibinfo
  {author} {\bibfnamefont {E.}~\bibnamefont {Seider}},\ and\ \bibinfo {author}
  {\bibfnamefont {L.}~\bibnamefont {Genzel}},\ }\href
  {https://doi.org/https://doi.org/10.1016/0921-4534(91)90771-P} {\bibfield
  {journal} {\bibinfo  {journal} {Physica C: Superconductivity}\ }\textbf
  {\bibinfo {volume} {183}},\ \bibinfo {pages} {99} (\bibinfo {year}
  {1991})}\BibitemShut {NoStop}%
\bibitem [{\citenamefont {Palmer}\ and\ \citenamefont
  {Tinkham}(1968)}]{PhysRev.165.588}%
  \BibitemOpen
  \bibfield  {author} {\bibinfo {author} {\bibfnamefont {L.~H.}\ \bibnamefont
  {Palmer}}\ and\ \bibinfo {author} {\bibfnamefont {M.}~\bibnamefont
  {Tinkham}},\ }\href {https://doi.org/10.1103/PhysRev.165.588} {\bibfield
  {journal} {\bibinfo  {journal} {Phys. Rev.}\ }\textbf {\bibinfo {volume}
  {165}},\ \bibinfo {pages} {588} (\bibinfo {year} {1968})}\BibitemShut
  {NoStop}%
\bibitem [{\citenamefont {Nambu}(1960)}]{PhysRev.117.648}%
  \BibitemOpen
  \bibfield  {author} {\bibinfo {author} {\bibfnamefont {Y.}~\bibnamefont
  {Nambu}},\ }\href {https://doi.org/10.1103/PhysRev.117.648} {\bibfield
  {journal} {\bibinfo  {journal} {Phys. Rev.}\ }\textbf {\bibinfo {volume}
  {117}},\ \bibinfo {pages} {648} (\bibinfo {year} {1960})}\BibitemShut
  {NoStop}%
\bibitem [{\citenamefont {Schrieffer}(1964)}]{Schrieffer}%
  \BibitemOpen
  \bibfield  {author} {\bibinfo {author} {\bibfnamefont {J.~F.}\ \bibnamefont
  {Schrieffer}},\ }\href@noop {} {\emph {\bibinfo {title} {{Theory of
  Superconductivity}}}}\ (\bibinfo  {publisher} {Benjamin, New York},\ \bibinfo
  {year} {1964})\BibitemShut {NoStop}%
\bibitem [{\citenamefont {Huang}\ and\ \citenamefont
  {Wang}(2023)}]{PhysRevB.108.224516}%
  \BibitemOpen
  \bibfield  {author} {\bibinfo {author} {\bibfnamefont {L.}~\bibnamefont
  {Huang}}\ and\ \bibinfo {author} {\bibfnamefont {J.}~\bibnamefont {Wang}},\
  }\href {https://doi.org/10.1103/PhysRevB.108.224516} {\bibfield  {journal}
  {\bibinfo  {journal} {Phys. Rev. B}\ }\textbf {\bibinfo {volume} {108}},\
  \bibinfo {pages} {224516} (\bibinfo {year} {2023})}\BibitemShut {NoStop}%
\bibitem [{\citenamefont {Watanabe}\ and\ \citenamefont
  {Watanabe}(2025{\natexlab{a}})}]{nonlineargauge1}%
  \BibitemOpen
  \bibfield  {author} {\bibinfo {author} {\bibfnamefont {S.}~\bibnamefont
  {Watanabe}}\ and\ \bibinfo {author} {\bibfnamefont {H.}~\bibnamefont
  {Watanabe}},\ }\href {https://doi.org/10.48550/arXiv.2410.18679} {\bibfield
  {journal} {\bibinfo  {journal} {arXiv: 2410.18679}\ } (\bibinfo {year}
  {2025}{\natexlab{a}})}\BibitemShut {NoStop}%
\bibitem [{\citenamefont {Watanabe}\ and\ \citenamefont
  {Watanabe}(2025{\natexlab{b}})}]{nonlineargauge}%
  \BibitemOpen
  \bibfield  {author} {\bibinfo {author} {\bibfnamefont {S.}~\bibnamefont
  {Watanabe}}\ and\ \bibinfo {author} {\bibfnamefont {H.}~\bibnamefont
  {Watanabe}},\ }\href {https://doi.org/10.48550/arXiv.2501.13722} {\bibfield
  {journal} {\bibinfo  {journal} {arXiv: 2501.13722}\ } (\bibinfo {year}
  {2025}{\natexlab{b}})}\BibitemShut {NoStop}%
\bibitem [{\citenamefont {Nakamura}\ \emph {et~al.}(2019)\citenamefont
  {Nakamura}, \citenamefont {Iida}, \citenamefont {Murotani}, \citenamefont
  {Matsunaga}, \citenamefont {Terai},\ and\ \citenamefont
  {Shimano}}]{PhysRevLett.122.257001}%
  \BibitemOpen
  \bibfield  {author} {\bibinfo {author} {\bibfnamefont {S.}~\bibnamefont
  {Nakamura}}, \bibinfo {author} {\bibfnamefont {Y.}~\bibnamefont {Iida}},
  \bibinfo {author} {\bibfnamefont {Y.}~\bibnamefont {Murotani}}, \bibinfo
  {author} {\bibfnamefont {R.}~\bibnamefont {Matsunaga}}, \bibinfo {author}
  {\bibfnamefont {H.}~\bibnamefont {Terai}},\ and\ \bibinfo {author}
  {\bibfnamefont {R.}~\bibnamefont {Shimano}},\ }\href
  {https://doi.org/10.1103/PhysRevLett.122.257001} {\bibfield  {journal}
  {\bibinfo  {journal} {Phys. Rev. Lett.}\ }\textbf {\bibinfo {volume} {122}},\
  \bibinfo {pages} {257001} (\bibinfo {year} {2019})}\BibitemShut {NoStop}%
\bibitem [{\citenamefont {Nakamura}\ \emph {et~al.}(2020)\citenamefont
  {Nakamura}, \citenamefont {Katsumi}, \citenamefont {Terai},\ and\
  \citenamefont {Shimano}}]{PhysRevLett.125.097004}%
  \BibitemOpen
  \bibfield  {author} {\bibinfo {author} {\bibfnamefont {S.}~\bibnamefont
  {Nakamura}}, \bibinfo {author} {\bibfnamefont {K.}~\bibnamefont {Katsumi}},
  \bibinfo {author} {\bibfnamefont {H.}~\bibnamefont {Terai}},\ and\ \bibinfo
  {author} {\bibfnamefont {R.}~\bibnamefont {Shimano}},\ }\href
  {https://doi.org/10.1103/PhysRevLett.125.097004} {\bibfield  {journal}
  {\bibinfo  {journal} {Phys. Rev. Lett.}\ }\textbf {\bibinfo {volume} {125}},\
  \bibinfo {pages} {097004} (\bibinfo {year} {2020})}\BibitemShut {NoStop}%
\bibitem [{\citenamefont {Du}\ \emph {et~al.}(2021)\citenamefont {Du},
  \citenamefont {Lu},\ and\ \citenamefont {Xie}}]{RefNonlineHallEff01}%
  \BibitemOpen
  \bibfield  {author} {\bibinfo {author} {\bibfnamefont {Z.~Z.}\ \bibnamefont
  {Du}}, \bibinfo {author} {\bibfnamefont {H.-Z.}\ \bibnamefont {Lu}},\ and\
  \bibinfo {author} {\bibfnamefont {X.~C.}\ \bibnamefont {Xie}},\ }\href
  {https://doi.org/10.1038/s42254-021-00359-6} {\bibfield  {journal} {\bibinfo
  {journal} {Nature Reviews Physics}\ }\textbf {\bibinfo {volume} {3}},\
  \bibinfo {pages} {744} (\bibinfo {year} {2021})}\BibitemShut {NoStop}%
\bibitem [{\citenamefont {Moll}\ and\ \citenamefont
  {Geshkenbein}(2023)}]{RefScDiods2}%
  \BibitemOpen
  \bibfield  {author} {\bibinfo {author} {\bibfnamefont {P.~J.~W.}\
  \bibnamefont {Moll}}\ and\ \bibinfo {author} {\bibfnamefont {V.~B.}\
  \bibnamefont {Geshkenbein}},\ }\href
  {https://doi.org/10.1038/s41567-023-02229-7} {\bibfield  {journal} {\bibinfo
  {journal} {Nature Physics}\ }\textbf {\bibinfo {volume} {19}},\ \bibinfo
  {pages} {1379} (\bibinfo {year} {2023})}\BibitemShut {NoStop}%
\bibitem [{\citenamefont {Strambini}\ \emph {et~al.}(2022)\citenamefont
  {Strambini}, \citenamefont {Spies}, \citenamefont {Ligato}, \citenamefont
  {Ili{\'c}}, \citenamefont {Rouco}, \citenamefont {Gonz{\'a}lez-Orellana},
  \citenamefont {Ilyn}, \citenamefont {Rogero}, \citenamefont {Bergeret},
  \citenamefont {Moodera}, \citenamefont {Virtanen}, \citenamefont
  {Heikkil{\"a}},\ and\ \citenamefont {Giazotto}}]{RefStrambini}%
  \BibitemOpen
  \bibfield  {author} {\bibinfo {author} {\bibfnamefont {E.}~\bibnamefont
  {Strambini}}, \bibinfo {author} {\bibfnamefont {M.}~\bibnamefont {Spies}},
  \bibinfo {author} {\bibfnamefont {N.}~\bibnamefont {Ligato}}, \bibinfo
  {author} {\bibfnamefont {S.}~\bibnamefont {Ili{\'c}}}, \bibinfo {author}
  {\bibfnamefont {M.}~\bibnamefont {Rouco}}, \bibinfo {author} {\bibfnamefont
  {C.}~\bibnamefont {Gonz{\'a}lez-Orellana}}, \bibinfo {author} {\bibfnamefont
  {M.}~\bibnamefont {Ilyn}}, \bibinfo {author} {\bibfnamefont {C.}~\bibnamefont
  {Rogero}}, \bibinfo {author} {\bibfnamefont {F.~S.}\ \bibnamefont
  {Bergeret}}, \bibinfo {author} {\bibfnamefont {J.~S.}\ \bibnamefont
  {Moodera}}, \bibinfo {author} {\bibfnamefont {P.}~\bibnamefont {Virtanen}},
  \bibinfo {author} {\bibfnamefont {T.~T.}\ \bibnamefont {Heikkil{\"a}}},\ and\
  \bibinfo {author} {\bibfnamefont {F.}~\bibnamefont {Giazotto}},\ }\href
  {https://doi.org/10.1038/s41467-022-29990-2} {\bibfield  {journal} {\bibinfo
  {journal} {Nature Communications}\ }\textbf {\bibinfo {volume} {13}},\
  \bibinfo {pages} {2431} (\bibinfo {year} {2022})}\BibitemShut {NoStop}%
\bibitem [{\citenamefont {Nadeem}\ \emph {et~al.}(2023)\citenamefont {Nadeem},
  \citenamefont {Fuhrer},\ and\ \citenamefont {Wang}}]{nadeem}%
  \BibitemOpen
  \bibfield  {author} {\bibinfo {author} {\bibfnamefont {M.}~\bibnamefont
  {Nadeem}}, \bibinfo {author} {\bibfnamefont {M.~S.}\ \bibnamefont {Fuhrer}},\
  and\ \bibinfo {author} {\bibfnamefont {X.}~\bibnamefont {Wang}},\ }\href
  {https://doi.org/10.1038/s42254-023-00632-w} {\bibfield  {journal} {\bibinfo
  {journal} {Nature Reviews Physics}\ }\textbf {\bibinfo {volume} {5}},\
  \bibinfo {pages} {558} (\bibinfo {year} {2023})}\BibitemShut {NoStop}%
\bibitem [{\citenamefont {Daido}\ \emph {et~al.}(2022)\citenamefont {Daido},
  \citenamefont {Ikeda},\ and\ \citenamefont
  {Yanase}}]{PhysRevLett.128.037001}%
  \BibitemOpen
  \bibfield  {author} {\bibinfo {author} {\bibfnamefont {A.}~\bibnamefont
  {Daido}}, \bibinfo {author} {\bibfnamefont {Y.}~\bibnamefont {Ikeda}},\ and\
  \bibinfo {author} {\bibfnamefont {Y.}~\bibnamefont {Yanase}},\ }\href
  {https://doi.org/10.1103/PhysRevLett.128.037001} {\bibfield  {journal}
  {\bibinfo  {journal} {Phys. Rev. Lett.}\ }\textbf {\bibinfo {volume} {128}},\
  \bibinfo {pages} {037001} (\bibinfo {year} {2022})}\BibitemShut {NoStop}%
\bibitem [{\citenamefont {Ili\ifmmode~\acute{c}\else \'{c}\fi{}}\ and\
  \citenamefont {Bergeret}(2022)}]{PhysRevLett.128.177001}%
  \BibitemOpen
  \bibfield  {author} {\bibinfo {author} {\bibfnamefont {S.}~\bibnamefont
  {Ili\ifmmode~\acute{c}\else \'{c}\fi{}}}\ and\ \bibinfo {author}
  {\bibfnamefont {F.~S.}\ \bibnamefont {Bergeret}},\ }\href
  {https://doi.org/10.1103/PhysRevLett.128.177001} {\bibfield  {journal}
  {\bibinfo  {journal} {Phys. Rev. Lett.}\ }\textbf {\bibinfo {volume} {128}},\
  \bibinfo {pages} {177001} (\bibinfo {year} {2022})}\BibitemShut {NoStop}%
\bibitem [{\citenamefont {He}\ \emph {et~al.}(2022)\citenamefont {He},
  \citenamefont {Tanaka},\ and\ \citenamefont {Nagaosa}}]{He_2022}%
  \BibitemOpen
  \bibfield  {author} {\bibinfo {author} {\bibfnamefont {J.~J.}\ \bibnamefont
  {He}}, \bibinfo {author} {\bibfnamefont {Y.}~\bibnamefont {Tanaka}},\ and\
  \bibinfo {author} {\bibfnamefont {N.}~\bibnamefont {Nagaosa}},\ }\href
  {https://doi.org/10.1088/1367-2630/ac6766} {\bibfield  {journal} {\bibinfo
  {journal} {New Journal of Physics}\ }\textbf {\bibinfo {volume} {24}},\
  \bibinfo {pages} {053014} (\bibinfo {year} {2022})}\BibitemShut {NoStop}%
\bibitem [{\citenamefont {Hu}\ \emph {et~al.}(2007)\citenamefont {Hu},
  \citenamefont {Wu},\ and\ \citenamefont {Dai}}]{PhysRevLett.99.067004}%
  \BibitemOpen
  \bibfield  {author} {\bibinfo {author} {\bibfnamefont {J.}~\bibnamefont
  {Hu}}, \bibinfo {author} {\bibfnamefont {C.}~\bibnamefont {Wu}},\ and\
  \bibinfo {author} {\bibfnamefont {X.}~\bibnamefont {Dai}},\ }\href
  {https://doi.org/10.1103/PhysRevLett.99.067004} {\bibfield  {journal}
  {\bibinfo  {journal} {Phys. Rev. Lett.}\ }\textbf {\bibinfo {volume} {99}},\
  \bibinfo {pages} {067004} (\bibinfo {year} {2007})}\BibitemShut {NoStop}%
\bibitem [{\citenamefont {Davydova}\ \emph {et~al.}(2022)\citenamefont
  {Davydova}, \citenamefont {Prembabu},\ and\ \citenamefont
  {Fu}}]{sciadv.abo0309}%
  \BibitemOpen
  \bibfield  {author} {\bibinfo {author} {\bibfnamefont {M.}~\bibnamefont
  {Davydova}}, \bibinfo {author} {\bibfnamefont {S.}~\bibnamefont {Prembabu}},\
  and\ \bibinfo {author} {\bibfnamefont {L.}~\bibnamefont {Fu}},\ }\href
  {https://doi.org/10.1126/sciadv.abo0309} {\bibfield  {journal} {\bibinfo
  {journal} {Science Advances}\ }\textbf {\bibinfo {volume} {8}},\ \bibinfo
  {pages} {eabo0309} (\bibinfo {year} {2022})}\BibitemShut {NoStop}%
\bibitem [{\citenamefont {Zhang}\ \emph {et~al.}(2022)\citenamefont {Zhang},
  \citenamefont {Gu}, \citenamefont {Li}, \citenamefont {Hu},\ and\
  \citenamefont {Jiang}}]{PhysRevX.12.041013}%
  \BibitemOpen
  \bibfield  {author} {\bibinfo {author} {\bibfnamefont {Y.}~\bibnamefont
  {Zhang}}, \bibinfo {author} {\bibfnamefont {Y.}~\bibnamefont {Gu}}, \bibinfo
  {author} {\bibfnamefont {P.}~\bibnamefont {Li}}, \bibinfo {author}
  {\bibfnamefont {J.}~\bibnamefont {Hu}},\ and\ \bibinfo {author}
  {\bibfnamefont {K.}~\bibnamefont {Jiang}},\ }\href
  {https://doi.org/10.1103/PhysRevX.12.041013} {\bibfield  {journal} {\bibinfo
  {journal} {Phys. Rev. X}\ }\textbf {\bibinfo {volume} {12}},\ \bibinfo
  {pages} {041013} (\bibinfo {year} {2022})}\BibitemShut {NoStop}%
\bibitem [{\citenamefont {Parafilo}\ \emph {et~al.}(2023)\citenamefont
  {Parafilo}, \citenamefont {Kovalev},\ and\ \citenamefont
  {Savenko}}]{PhysRevB.108.L180509}%
  \BibitemOpen
  \bibfield  {author} {\bibinfo {author} {\bibfnamefont {A.~V.}\ \bibnamefont
  {Parafilo}}, \bibinfo {author} {\bibfnamefont {V.~M.}\ \bibnamefont
  {Kovalev}},\ and\ \bibinfo {author} {\bibfnamefont {I.~G.}\ \bibnamefont
  {Savenko}},\ }\href {https://doi.org/10.1103/PhysRevB.108.L180509} {\bibfield
   {journal} {\bibinfo  {journal} {Phys. Rev. B}\ }\textbf {\bibinfo {volume}
  {108}},\ \bibinfo {pages} {L180509} (\bibinfo {year} {2023})}\BibitemShut
  {NoStop}%
\bibitem [{\citenamefont {Sonowal}\ \emph {et~al.}(2024)\citenamefont
  {Sonowal}, \citenamefont {Parafilo}, \citenamefont {Kovalev},\ and\
  \citenamefont {Savenko}}]{PhysRevB.110.205413}%
  \BibitemOpen
  \bibfield  {author} {\bibinfo {author} {\bibfnamefont {K.}~\bibnamefont
  {Sonowal}}, \bibinfo {author} {\bibfnamefont {A.~V.}\ \bibnamefont
  {Parafilo}}, \bibinfo {author} {\bibfnamefont {V.~M.}\ \bibnamefont
  {Kovalev}},\ and\ \bibinfo {author} {\bibfnamefont {I.~G.}\ \bibnamefont
  {Savenko}},\ }\href {https://doi.org/10.1103/PhysRevB.110.205413} {\bibfield
  {journal} {\bibinfo  {journal} {Phys. Rev. B}\ }\textbf {\bibinfo {volume}
  {110}},\ \bibinfo {pages} {205413} (\bibinfo {year} {2024})}\BibitemShut
  {NoStop}%
\bibitem [{\citenamefont {Parafilo}\ \emph {et~al.}(2024)\citenamefont
  {Parafilo}, \citenamefont {Sun}, \citenamefont {Sonowal}, \citenamefont
  {Kovalev},\ and\ \citenamefont {Savenko}}]{Parafilo_2025}%
  \BibitemOpen
  \bibfield  {author} {\bibinfo {author} {\bibfnamefont {A.~V.}\ \bibnamefont
  {Parafilo}}, \bibinfo {author} {\bibfnamefont {M.}~\bibnamefont {Sun}},
  \bibinfo {author} {\bibfnamefont {K.}~\bibnamefont {Sonowal}}, \bibinfo
  {author} {\bibfnamefont {V.~M.}\ \bibnamefont {Kovalev}},\ and\ \bibinfo
  {author} {\bibfnamefont {I.~G.}\ \bibnamefont {Savenko}},\ }\href
  {https://doi.org/10.1088/2053-1583/ad9596} {\bibfield  {journal} {\bibinfo
  {journal} {2D Materials}\ }\textbf {\bibinfo {volume} {12}},\ \bibinfo
  {pages} {011001} (\bibinfo {year} {2024})}\BibitemShut {NoStop}%
\bibitem [{\citenamefont {Gal'perin}\ \emph {et~al.}(1978)\citenamefont
  {Gal'perin}, \citenamefont {Gurevich},\ and\ \citenamefont
  {Kozub}}]{PhysRevB.18.5116}%
  \BibitemOpen
  \bibfield  {author} {\bibinfo {author} {\bibfnamefont {Y.~M.}\ \bibnamefont
  {Gal'perin}}, \bibinfo {author} {\bibfnamefont {V.~L.}\ \bibnamefont
  {Gurevich}},\ and\ \bibinfo {author} {\bibfnamefont {V.~I.}\ \bibnamefont
  {Kozub}},\ }\href {https://doi.org/10.1103/PhysRevB.18.5116} {\bibfield
  {journal} {\bibinfo  {journal} {Phys. Rev. B}\ }\textbf {\bibinfo {volume}
  {18}},\ \bibinfo {pages} {5116} (\bibinfo {year} {1978})}\BibitemShut
  {NoStop}%
\bibitem [{\citenamefont {Van~Harlingen}\ \emph {et~al.}(1980)\citenamefont
  {Van~Harlingen}, \citenamefont {Heidel},\ and\ \citenamefont
  {Garland}}]{PhysRevB.21.1842}%
  \BibitemOpen
  \bibfield  {author} {\bibinfo {author} {\bibfnamefont {D.~J.}\ \bibnamefont
  {Van~Harlingen}}, \bibinfo {author} {\bibfnamefont {D.~F.}\ \bibnamefont
  {Heidel}},\ and\ \bibinfo {author} {\bibfnamefont {J.~C.}\ \bibnamefont
  {Garland}},\ }\href {https://doi.org/10.1103/PhysRevB.21.1842} {\bibfield
  {journal} {\bibinfo  {journal} {Phys. Rev. B}\ }\textbf {\bibinfo {volume}
  {21}},\ \bibinfo {pages} {1842} (\bibinfo {year} {1980})}\BibitemShut
  {NoStop}%
\bibitem [{\citenamefont {Galperin}\ \emph {et~al.}(2002)\citenamefont
  {Galperin}, \citenamefont {Gurevich}, \citenamefont {Kozub},\ and\
  \citenamefont {Shelankov}}]{PhysRevB.65.064531}%
  \BibitemOpen
  \bibfield  {author} {\bibinfo {author} {\bibfnamefont {Y.~M.}\ \bibnamefont
  {Galperin}}, \bibinfo {author} {\bibfnamefont {V.~L.}\ \bibnamefont
  {Gurevich}}, \bibinfo {author} {\bibfnamefont {V.~I.}\ \bibnamefont
  {Kozub}},\ and\ \bibinfo {author} {\bibfnamefont {A.~L.}\ \bibnamefont
  {Shelankov}},\ }\href {https://doi.org/10.1103/PhysRevB.65.064531} {\bibfield
   {journal} {\bibinfo  {journal} {Phys. Rev. B}\ }\textbf {\bibinfo {volume}
  {65}},\ \bibinfo {pages} {064531} (\bibinfo {year} {2002})}\BibitemShut
  {NoStop}%
\bibitem [{\citenamefont {Smith}\ \emph
  {et~al.}(2020{\natexlab{a}})\citenamefont {Smith}, \citenamefont {Andreev},\
  and\ \citenamefont {Spivak}}]{PhysRevB.101.134508}%
  \BibitemOpen
  \bibfield  {author} {\bibinfo {author} {\bibfnamefont {M.}~\bibnamefont
  {Smith}}, \bibinfo {author} {\bibfnamefont {A.~V.}\ \bibnamefont {Andreev}},\
  and\ \bibinfo {author} {\bibfnamefont {B.~Z.}\ \bibnamefont {Spivak}},\
  }\href {https://doi.org/10.1103/PhysRevB.101.134508} {\bibfield  {journal}
  {\bibinfo  {journal} {Phys. Rev. B}\ }\textbf {\bibinfo {volume} {101}},\
  \bibinfo {pages} {134508} (\bibinfo {year} {2020}{\natexlab{a}})}\BibitemShut
  {NoStop}%
\bibitem [{\citenamefont {Smith}\ \emph
  {et~al.}(2020{\natexlab{b}})\citenamefont {Smith}, \citenamefont {Andreev},\
  and\ \citenamefont {Spivak}}]{SMITH2020168105}%
  \BibitemOpen
  \bibfield  {author} {\bibinfo {author} {\bibfnamefont {M.}~\bibnamefont
  {Smith}}, \bibinfo {author} {\bibfnamefont {A.}~\bibnamefont {Andreev}},\
  and\ \bibinfo {author} {\bibfnamefont {B.}~\bibnamefont {Spivak}},\ }\href
  {https://doi.org/https://doi.org/10.1016/j.aop.2020.168105} {\bibfield
  {journal} {\bibinfo  {journal} {Annals of Physics}\ }\textbf {\bibinfo
  {volume} {417}},\ \bibinfo {pages} {168105} (\bibinfo {year}
  {2020}{\natexlab{b}})}\BibitemShut {NoStop}%
\bibitem [{\citenamefont {Ovchinnikov}\ and\ \citenamefont
  {lsaakyan}(1978)}]{Ovchinnlsaakyan}%
  \BibitemOpen
  \bibfield  {author} {\bibinfo {author} {\bibfnamefont {Y.~N.}\ \bibnamefont
  {Ovchinnikov}}\ and\ \bibinfo {author} {\bibfnamefont {A.~R.}\ \bibnamefont
  {lsaakyan}},\ }\href@noop {} {\bibfield  {journal} {\bibinfo  {journal} {Zh.
  Eksp. Teor. Fiz}\ }\textbf {\bibinfo {volume} {74}},\ \bibinfo {pages} {178}
  (\bibinfo {year} {1978})}\BibitemShut {NoStop}%
\bibitem [{\citenamefont {Reizer}(1998)}]{PhysRevB.57.1147}%
  \BibitemOpen
  \bibfield  {author} {\bibinfo {author} {\bibfnamefont {M.}~\bibnamefont
  {Reizer}},\ }\href {https://doi.org/10.1103/PhysRevB.57.1147} {\bibfield
  {journal} {\bibinfo  {journal} {Phys. Rev. B}\ }\textbf {\bibinfo {volume}
  {57}},\ \bibinfo {pages} {1147} (\bibinfo {year} {1998})}\BibitemShut
  {NoStop}%
\bibitem [{\citenamefont {Reizer}(2000)}]{PhysRevB.61.7108}%
  \BibitemOpen
  \bibfield  {author} {\bibinfo {author} {\bibfnamefont {M.}~\bibnamefont
  {Reizer}},\ }\href {https://doi.org/10.1103/PhysRevB.61.7108} {\bibfield
  {journal} {\bibinfo  {journal} {Phys. Rev. B}\ }\textbf {\bibinfo {volume}
  {61}},\ \bibinfo {pages} {7108} (\bibinfo {year} {2000})}\BibitemShut
  {NoStop}%
\bibitem [{\citenamefont {Peskin}\ and\ \citenamefont
  {Schroeder}(1995)}]{Peskin}%
  \BibitemOpen
  \bibfield  {author} {\bibinfo {author} {\bibfnamefont {M.}~\bibnamefont
  {Peskin}}\ and\ \bibinfo {author} {\bibfnamefont {D.}~\bibnamefont
  {Schroeder}},\ }\href@noop {} {\emph {\bibinfo {title} {{An Introduction to
  Quantum Field Theory}}}}\ (\bibinfo  {publisher} {Westview Press, Colorado},\
  \bibinfo {year} {1995})\BibitemShut {NoStop}%
\bibitem [{\citenamefont {Sidorova}\ \emph {et~al.}(2020)\citenamefont
  {Sidorova}, \citenamefont {Semenov}, \citenamefont {H\"ubers}, \citenamefont
  {Ilin}, \citenamefont {Siegel}, \citenamefont {Charaev}, \citenamefont
  {Moshkova}, \citenamefont {Kaurova}, \citenamefont {Goltsman}, \citenamefont
  {Zhang},\ and\ \citenamefont {Schilling}}]{PhysRevB.102.054501}%
  \BibitemOpen
  \bibfield  {author} {\bibinfo {author} {\bibfnamefont {M.}~\bibnamefont
  {Sidorova}}, \bibinfo {author} {\bibfnamefont {A.}~\bibnamefont {Semenov}},
  \bibinfo {author} {\bibfnamefont {H.-W.}\ \bibnamefont {H\"ubers}}, \bibinfo
  {author} {\bibfnamefont {K.}~\bibnamefont {Ilin}}, \bibinfo {author}
  {\bibfnamefont {M.}~\bibnamefont {Siegel}}, \bibinfo {author} {\bibfnamefont
  {I.}~\bibnamefont {Charaev}}, \bibinfo {author} {\bibfnamefont
  {M.}~\bibnamefont {Moshkova}}, \bibinfo {author} {\bibfnamefont
  {N.}~\bibnamefont {Kaurova}}, \bibinfo {author} {\bibfnamefont {G.~N.}\
  \bibnamefont {Goltsman}}, \bibinfo {author} {\bibfnamefont {X.}~\bibnamefont
  {Zhang}},\ and\ \bibinfo {author} {\bibfnamefont {A.}~\bibnamefont
  {Schilling}},\ }\href {https://doi.org/10.1103/PhysRevB.102.054501}
  {\bibfield  {journal} {\bibinfo  {journal} {Phys. Rev. B}\ }\textbf {\bibinfo
  {volume} {102}},\ \bibinfo {pages} {054501} (\bibinfo {year}
  {2020})}\BibitemShut {NoStop}%
\end{thebibliography}%
\bibliographystyle{apsrev4-2}


\end{document}